\documentclass[twocolumn,trackchanges]{aastex631}

\shorttitle{HAWC~J1844-034 Analysis}
\shortauthors{THE HAWC COLLABORATION}

\graphicspath{{./}{figures/}}

\begin{document}

\title{HAWC Study of Very-High-Energy $\gamma$-ray Spectrum of HAWC~J1844-034}

\correspondingauthor{C.D.~Rho}
\email{cdr397@skku.edu}
\correspondingauthor{Y.~Son}
\email{youngwan.son@cern.ch}

\received{2023.04.25}
\revised{2023.07.17}
\accepted{2023.07.19}
\submitjournal{ApJ}

\author[0000-0003-0197-5646]{A.~Albert}
\affiliation{Physics Division, Los Alamos National Laboratory, Los Alamos, NM, USA}

\author[0000-0001-8310-4486]{C.~Alvarez}
\affiliation{Universidad Aut\'{o}noma de Chiapas, Tuxtla Guti\'{e}rrez, Chiapas, M\'{e}xico}

\author[0000-0002-4020-4142]{D.~Avila Rojas}
\affiliation{Instituto de F\'{i}sica, Universidad Nacional Aut\'{o}noma de M\'{e}xico, Ciudad de Mexico, Mexico}

\author[0000-0002-2084-5049]{H.A.~Ayala Solares}
\affiliation{Department of Physics, Pennsylvania State University, University Park, PA, USA}

\author[0000-0002-5529-6780]{R.~Babu}
\affiliation{Department of Physics, Michigan Technological University, Houghton, MI, USA}

\author[0000-0003-3207-105X]{E.~Belmont-Moreno}
\affiliation{Instituto de F\'{i}sica, Universidad Nacional Aut\'{o}noma de M\'{e}xico, Ciudad de Mexico, Mexico}

\author[0000-0003-0268-5122]{M.~Breuhaus}
\affiliation{Max-Planck Institute for Nuclear Physics, 69117 Heidelberg, Germany}

\author[0000-0003-2158-2292]{T.~Capistr\'{a}n}
\affiliation{Instituto de Astronom\'{i}a, Universidad Nacional Aut\'{o}noma de M\'{e}xico, Ciudad de Mexico, Mexico}

\author[0000-0002-8553-3302]{A.~Carrami\~{n}ana}
\affiliation{Instituto Nacional de Astrof\'{i}sica, \'{O}ptica y Electr\'{o}nica, Puebla, Mexico}

\author[0000-0002-6144-9122]{S.~Casanova}
\affiliation{Institute of Nuclear Physics Polish Academy of Sciences, PL-31342 IFJ-PAN, Krakow, Poland}

\author[0000-0002-1132-871X]{J.~Cotzomi}
\affiliation{Facultad de Ciencias F\'{i}sico Matem\'{a}ticas, Benem\'{e}rita Universidad Aut\'{o}noma de Puebla, Puebla, Mexico}

\author[0000-0002-7747-754X]{S.~Couti\~{n}o de Le\'{o}n}
\affiliation{Department of Physics, University of Wisconsin-Madison, Madison, WI, USA}

\author[0000-0001-9643-4134]{E.~De la Fuente}
\affiliation{Departamento de F\'{i}sica, Centro Universitario de Ciencias Exactase Ingenierias, Universidad de Guadalajara, Guadalajara, Mexico}

\author[0000-0002-2672-4141]{D.~Depaoli}
\affiliation{Max-Planck Institute for Nuclear Physics, 69117 Heidelberg, Germany}

\author[0000-0001-8487-0836]{R.~Diaz Hernandez}
\affiliation{Instituto Nacional de Astrof\'{i}sica, \'{O}ptica y Electr\'{o}nica, Puebla, Mexico}

\author[0000-0001-8451-7450]{B.L.~Dingus}
\affiliation{Physics Division, Los Alamos National Laboratory, Los Alamos, NM, USA}

\author[0000-0002-2987-9691]{M.A.~DuVernois}
\affiliation{Department of Physics, University of Wisconsin-Madison, Madison, WI, USA}

\author[0000-0003-2169-0306]{M.~Durocher}
\affiliation{Physics Division, Los Alamos National Laboratory, Los Alamos, NM, USA}

\author[0000-0001-5737-1820]{K.~Engel}
\affiliation{Department of Physics, University of Maryland, College Park, MD, USA}

\author[0000-0001-7074-1726]{C.~Espinoza}
\affiliation{Instituto de F\'{i}sica, Universidad Nacional Aut\'{o}noma de M\'{e}xico, Ciudad de Mexico, Mexico}

\author[0000-0002-8246-4751]{K.L.~Fan}
\affiliation{Department of Physics, University of Maryland, College Park, MD, USA}

\author[0000-0002-5387-8138]{K.~Fang}
\affiliation{Department of Physics, University of Wisconsin-Madison, Madison, WI, USA}

\author[0000-0002-0173-6453]{N.~Fraija}
\affiliation{Instituto de Astronom\'{i}a, Universidad Nacional Aut\'{o}noma de M\'{e}xico, Ciudad de Mexico, Mexico}

\author[0000-0002-4188-5584]{J.A.~Garc\'{i}a-Gonz\'{a}lez}
\affiliation{Tecnologico de Monterrey, Escuela de Ingenier\'{i}a y Ciencias, Ave. Eugenio Garza Sada 2501, Monterrey, N.L., Mexico, 64849}

\author[0000-0002-5209-5641]{M.M.~Gonz\'{a}lez}
\affiliation{Instituto de Astronom\'{i}a, Universidad Nacional Aut\'{o}noma de M\'{e}xico, Ciudad de Mexico, Mexico}

\author[0000-0002-9790-1299]{J.A.~Goodman}
\affiliation{Department of Physics, University of Maryland, College Park, MD, USA}

\author{S.~Groetsch}
\affiliation{Department of Physics, Michigan Technological University, Houghton, MI, USA}

\author[0000-0001-9844-2648]{J.P.~Harding}
\affiliation{Physics Division, Los Alamos National Laboratory, Los Alamos, NM, USA}

\author{I.~Herzog}
\affiliation{Department of Physics and Astronomy, Michigan State University, East Lansing, MI, USA}

\author[0000-0002-1031-7760]{J.~Hinton}
\affiliation{Max-Planck Institute for Nuclear Physics, 69117 Heidelberg, Germany}

\author[0000-0002-5447-1786]{D.~Huang}
\affiliation{Department of Physics, Michigan Technological University, Houghton, MI, USA}

\author[0000-0002-5527-7141]{F.~Hueyotl-Zahuantitla}
\affiliation{Universidad Aut\'{o}noma de Chiapas, Tuxtla Guti\'{e}rrez, Chiapas, M\'{e}xico}

\author[0000-0002-1432-7771]{T.B.~Humensky}
\affiliation{NASA Goddard Space Flight Center, Greenbelt, MD 20771, USA}

\author[0000-0002-3302-7897]{P.~H\"{u}ntemeyer}
\affiliation{Department of Physics, Michigan Technological University, Houghton, MI, USA}

\author[0000-0003-4467-3621]{V.~Joshi}
\affiliation{Erlangen Centre for Astroparticle Physics, Friedrich-Alexander-Universit\"{a}t Erlangen-N\"{u}rnberg, Erlangen, Germany}

\author{S.~Kaufmann}
\affiliation{Universidad Politecnica de Pachuca, Pachuca, Hgo, Mexico}

\author[0000-0002-2153-1519]{J.~Lee}
\affiliation{University of Seoul, Seoul, Rep. of Korea}

\author[0000-0001-5516-4975]{H.~Le\'{o}n Vargas}
\affiliation{Instituto de F\'{i}sica, Universidad Nacional Aut\'{o}noma de M\'{e}xico, Ciudad de Mexico, Mexico}

\author[0000-0001-8825-3624]{A.L.~Longinotti}
\affiliation{Instituto de Astronom\'{i}a, Universidad Nacional Aut\'{o}noma de M\'{e}xico, Ciudad de Mexico, Mexico}

\author[0000-0003-2810-4867]{G.~Luis-Raya}
\affiliation{Universidad Politecnica de Pachuca, Pachuca, Hgo, Mexico}

\author[0000-0001-8088-400X]{K.~Malone}
\affiliation{Space Science and Applications Group, Los Alamos National Laboratory, Los Alamos, NM, USA}

\author[0000-0001-9052-856X]{O.~Martinez}
\affiliation{Facultad de Ciencias F\'{i}sico Matem\'{a}ticas, Benem\'{e}rita Universidad Aut\'{o}noma de Puebla, Puebla, Mexico}

\author[0000-0002-2824-3544]{J.~Mart\'{i}nez-Castro}
\affiliation{Centro de Investigaci\'{o}n en Computaci\'{o}n, Instituto Polit\'{e}cnico Nacional, Ciudad de M\'{e}xico, M\'{e}xico}

\author[0000-0002-2610-863X]{J.A.~Matthews}
\affiliation{Dept of Physics and Astronomy, University of New Mexico, Albuquerque, NM, USA}

\author[0000-0002-8390-9011]{P.~Miranda-Romagnoli}
\affiliation{Universidad Aut\'{o}noma del Estado de Hidalgo, Pachuca, Mexico}

\author[0000-0001-9361-0147]{J.A.~Morales-Soto}
\affiliation{Universidad Michoacana de San Nicol\'{a}s de Hidalgo, Morelia, Mexico}

\author[0000-0002-1114-2640]{E.~Moreno}
\affiliation{Facultad de Ciencias F\'{i}sico Matem\'{a}ticas, Benem\'{e}rita Universidad Aut\'{o}noma de Puebla, Puebla, Mexico}

\author[0000-0002-7675-4656]{M.~Mostaf\'{a}}
\affiliation{Department of Physics, Pennsylvania State University, University Park, PA, USA}

\author[0000-0003-1059-8731]{L.~Nellen}
\affiliation{Instituto de Ciencias Nucleares, Universidad Nacional Aut\'{o}noma de Mexico, Ciudad de Mexico, Mexico}

\author[0000-0001-7099-108X]{R.~Noriega-Papaqui}
\affiliation{Universidad Aut\'{o}noma del Estado de Hidalgo, Pachuca, Mexico}

\author[0000-0002-9105-0518]{L.~Olivera-Nieto}
\affiliation{Max-Planck Institute for Nuclear Physics, 69117 Heidelberg, Germany}

\author[0000-0002-5448-7577]{N.~Omodei}
\affiliation{Department of Physics, Stanford University: Stanford, CA 94305–4060, USA}

\author[0000-0001-5998-4938]{E.G.~P\'{e}rez-P\'{e}rez}
\affiliation{Universidad Politecnica de Pachuca, Pachuca, Hgo, Mexico}

\author[0000-0002-6524-9769]{C.D.~Rho}
\affiliation{Department of Physics, Sungkyunkwan University, Suwon 16419, South Korea}

\author[0000-0003-1327-0838]{D.~Rosa-Gonz\'{a}lez}
\affiliation{Instituto Nacional de Astrof\'{i}sica, \'{O}ptica y Electr\'{o}nica, Puebla, Mexico}

\author[0000-0001-6939-7825]{E.~Ruiz-Velasco}
\affiliation{Max-Planck Institute for Nuclear Physics, 69117 Heidelberg, Germany}

\author[0000-0003-4556-7302]{H.~Salazar}
\affiliation{Facultad de Ciencias F\'{i}sico Matem\'{a}ticas, Benem\'{e}rita Universidad Aut\'{o}noma de Puebla, Puebla, Mexico}

\author[0000-0002-9312-9684]{D.~Salazar-Gallegos}
\affiliation{Department of Physics and Astronomy, Michigan State University, East Lansing, MI, USA}

\author[0000-0001-6079-2722]{A.~Sandoval}
\affiliation{Instituto de F\'{i}sica, Universidad Nacional Aut\'{o}noma de M\'{e}xico, Ciudad de Mexico, Mexico}

\author[0000-0001-8644-4734]{M.~Schneider}
\affiliation{Department of Physics, University of Maryland, College Park, MD, USA}

\author{J.~Serna-Franco}
\affiliation{Instituto de F\'{i}sica, Universidad Nacional Aut\'{o}noma de M\'{e}xico, Ciudad de Mexico, Mexico}

\author[0000-0002-1012-0431]{A.J.~Smith}
\affiliation{Department of Physics, University of Maryland, College Park, MD, USA}

\author[0000-0002-7214-8480]{Y.~Son}
\affiliation{University of Seoul, Seoul, Rep. of Korea}

\author[0000-0002-1492-0380]{R.W.~Springer}
\affiliation{Department of Physics and Astronomy, University of Utah, Salt Lake City, UT, USA}

\author[0000-0002-9074-0584]{O.~Tibolla}
\affiliation{Universidad Politecnica de Pachuca, Pachuca, Hgo, Mexico}

\author[0000-0001-9725-1479]{K.~Tollefson}
\affiliation{Department of Physics and Astronomy, Michigan State University, East Lansing, MI, USA}

\author[0000-0002-1689-3945]{I.~Torres}
\affiliation{Instituto Nacional de Astrof\'{i}sica, \'{O}ptica y Electr\'{o}nica, Puebla, Mexico}

\author[0000-0002-7102-3352]{R.~Torres-Escobedo}
\affiliation{Tsung-Dao Lee Institute \& School of Physics and Astronomy, Shanghai Jiao Tong University, Shanghai, China}

\author[0000-0003-1068-6707]{R.~Turner}
\affiliation{Department of Physics, Michigan Technological University, Houghton, MI, USA}

\author[0000-0002-2748-2527]{F.~Ure\~{n}a-Mena}
\affiliation{Instituto Nacional de Astrof\'{i}sica, \'{O}ptica y Electr\'{o}nica, Puebla, Mexico}

\author[0000-0003-0715-7513]{E.~Varela}
\affiliation{Facultad de Ciencias F\'{i}sico Matem\'{a}ticas, Benem\'{e}rita Universidad Aut\'{o}noma de Puebla, Puebla, Mexico}

\author[0000-0001-6876-2800]{L.~Villase\~{n}or}
\affiliation{Facultad de Ciencias F\'{i}sico Matem\'{a}ticas, Benem\'{e}rita Universidad Aut\'{o}noma de Puebla, Puebla, Mexico}

\author[0000-0001-6798-353X]{X.~Wang}
\affiliation{Department of Physics, Michigan Technological University, Houghton, MI, USA}

\author[0000-0003-2141-3413]{I.J.~Watson}
\affiliation{University of Seoul, Seoul, Rep. of Korea}

\author[0000-0002-6623-0277]{E.~Willox}
\affiliation{Department of Physics, University of Maryland, College Park, MD, USA}

\author[0000-0003-0513-3841]{H.~Zhou}
\affiliation{Tsung-Dao Lee Institute \& School of Physics and Astronomy, Shanghai Jiao Tong University, Shanghai, China}

\collaboration{76}{(HAWC collaboration)}

\begin{abstract}

Recently, the region surrounding eHWC~J1842-035 has been studied extensively by gamma-ray observatories due to its extended emission reaching up to a few hundred TeV and potential as a hadronic accelerator.
In this work, we use 1,910~days of cumulative data from the High Altitude Water Cherenkov (HAWC) observatory to carry out a dedicated systematic source search of the eHWC~J1842-035 region. During the search we have found three sources in the region, namely, HAWC~J1844-034, HAWC~J1843-032, and HAWC~J1846-025. We have identified HAWC~J1844-034 as the extended source that emits photons with energies up to 175~TeV. 
We compute the spectrum for HAWC J1844-034 and by comparing with the observational results from other experiments, we have identified HESS~J1843-033, LHAASO~J1843-0338, and TASG~J1844-038 as very-high-energy gamma-ray sources with a matching origin. 
Also, we present and use the multi-wavelength data to fit the hadronic and leptonic particle spectra. {We have identified four pulsar candidates in the nearby region from which PSR~J1844-0346 is found to be the most likely candidate due to its proximity to HAWC~J1844-034 and the computed energy budget. We have also found SNR~G28.6-0.1 as a potential counterpart source of HAWC~J1844-034 for which both leptonic and hadronic scenarios are feasible.} 
\end{abstract}

\section{Introduction} \label{sec:intro}

The origin of cosmic rays has been one of the primary questions the astrophysical community has been asking for over a century. Although there has been a lot of advancement in observational techniques and technology, our understanding of cosmic rays is far from complete. 
Cosmic ray sources cannot be traced directly due to deflections of cosmic ray particles, but $\gamma$-ray photons of the hadronic origin can be used to indirectly search for cosmic ray sources. However, $\gamma$-ray photons can also be produced from leptonic processes, hence a rigorous analysis is required to discern between hadronic-induced $\gamma$-ray sources and leptonic-induced $\gamma$-ray sources.

One particular feature of the cosmic ray spectrum is a break at around a few peta-electronvolts (PeVs) that is known as the ``knee'' \citep{Knee}.
Cosmic rays with energies less than the knee are believed to be accelerated by Galactic objects. 
These sources that can accelerate particles up to a PeV energy scale without a break are known as PeVatrons \citep{G106}.

If a Galactic gamma-ray source has been identified as a cosmic ray accelerator, we can study its spectrum to look for potential traces of a PeVatron. 
The proton spectrum of a PeVatron should cut off after a few PeV, leading to a hard $\gamma$-ray spectrum below the cutoff region corresponding to $\sim$100~TeV \citep{PeVatronSignature}. Therefore, the spectrum of $\gamma$-rays near the 100~TeV energy range can be useful for identifying PeVatron candidates \citep{PeVatron}.

eHWC~J1842-035 is one of the nine sources that was observed at above 56~TeV by the High Altitude Water Cherenkov (HAWC) Observatory in its ``High-Energy Catalog'' \citep{eHWC}. This result was later reinforced by the Large High Altitude Air Shower Observatory (LHAASO) in 2021 \citep{UHELHAASO}. LHAASO observed a bright extended source at a similar location of $0\fdg14$ from eHWC~J1842-035. LHAASO reported the $\gamma$-ray flux at 100~TeV. Also, in 2022 the Tibet AS-Gamma Experiment published observational results on an extended source, TASG~J1844-038, with a spectrum reaching up to 100~TeV \citep{Tibet}. The observations by the three different multi-TeV $\gamma$-ray observatories are highly likely to be originating from the same source.

In this work, we report the updated HAWC results on this bright extended source (to be referred to as HAWC~J1844-034) obtained from a rigorous multi-source modeling of the region (Section~\ref{sec:Results}). In Section~\ref{sec:Discussion}, we discuss the spectral studies on the results as well as the origin of the observed emission and systematic studies, providing evidence of counterpart sources in the region from other experiments.

\section{Instrument and Data Analysis} \label{sec:Instrument}

\subsection{HAWC}

HAWC is a ground-based particle sampling array located on the mountain Sierra Negra, Mexico at an altitude of 4,100~m. HAWC is designed to observe very-high-energy (VHE) $\gamma$-ray photons by detecting Cherenkov radiation produced from electromagnetic air showers as they propagate through the water Cherenkov detectors (WCDs). The main array of HAWC, composed of 300 WCDs, cover a geometrical area of 22,000 m${}^{2}$ {\citep{HAWCNIMA}}. Each WCD consists of a bladder containing {180,000~l} of purified water that is protected by a metal casing with a roof on top. A WCD has a height of {5.4~m} and a diameter of 7.3~m. Four photomultiplier tubes (PMTs) are located at the bottom of each WCD to collect the Cherenkov light emitted by air shower particles. For more detailed information on the design and operation of HAWC, refer to {\citet{HAWCCrab, HAWCNIMA}}. 

The gamma-ray events collected by HAWC are categorized into analysis bins based on the fraction of the PMTs used in the event reconstruction ($f_{\mathrm{hit}}$) and the estimated energies of the particular event \citep{HAWCCrab, HAWCEE}. We call these the ``ground parameter'' (GP) and ``neural network'' (NN) energy estimators.

The HAWC High-Energy Catalog released in 2020 used the GP dataset that contains 1,039~days of cumulative data \citep{eHWC}. 
In this work, we have used $1,910$~days of the GP energy estimator data above 17.8~TeV and $f_{\mathrm{hit}}$ above 24.5~\% that are within our region of interest (ROI) to compute the most sensitive results on the high energy spectrum of HAWC~J1844-034. {The energy resolution for the GP energy estimator is 50\% at 20~TeV and 34\% at 100~TeV \citep{HAWCEE}. The angular resolution range for the GP energy estimator bins used in the analysis is $0\fdg1 - 0\fdg6$ \citep{HAWCEE}.} Our ROI is a rectangular region defined as $28\degr < l < 30\degr$ and $-5\degr < b < 5\degr$, where $l$ and $b$ are the Galactic longitude and latitude, respectively. {The ROI is presented as the brighter region in Figure~\ref{fig:HAWCmap}, which has been chosen to include the extended emission.}

The multi-mission maximum likelihood (3ML)~\footnote{3ML; \url{https://github.com/threeML/threeML}} framework and HAWC accelerated likelihood (HAL)~\footnote{HAL; \url{https://github.com/threeML/hawc_hal}} plugin \citep{threeML, HAL} are used to simultaneously fit multiple source models and their relevant free parameters, based on maximum likelihood estimations. Spectral and morphological models mentioned in this work can be found in \textit{astromodels}~\footnote{\textit{astromodels}; \url{https://github.com/threeML/astromodels}}.

In our analysis, a test statistic (TS) is used to evaluate a pre-trial statistical significance of a fitted model in a region with a predetermined number of free parameters. 
TS provides a statistical measure of how well an alternative hypothesis performs over a null hypothesis. TS values can be used to compare between different nested models.
TS is defined as below:

\begin{equation}
   \mathrm{TS} = 2\mathrm{ln}\left(\frac{L_{1}}{L_{0}}\right),
\end{equation}
where $L_{1}$ is the likelihood for an alternative hypothesis and $L_{0}$ is the likelihood for a null hypothesis. To compare two different alternative hypotheses, $\Delta \mathrm{TS} = \mathrm{TS}_{2} - \mathrm{TS}_{1} = 2\mathrm{ln}\left(L_{2}/L_{1}\right)$ can be used, where $L_{2}$ is the likelihood for a different alternative hypothesis \citep{HAWCCrab}. Using Wilks' theorem \citep{Wilks}, a pre-trial significance can be computed with $\sigma \simeq \sqrt{\mathrm{TS}}$, {which is used to generate HAWC significance maps} \citep{2HWC}.

In order to find a statistically optimal list of sources in the ROI, a systematic approach developed by the Fermi Large Area Telescope (Fermi-LAT) Collaboration is used \citep{SystematicFermi}. 
For the source search, we recursively add a point source to the model in our ROI until the improvement of the overall TS becomes less than 16. Then, we test an extended source model for each of the found point sources. Starting from the brightest source, we accept an extended source model if $\Delta\mathrm{TS} > 16$.

During this process, we assumed a power-law spectral model for all of the sources involved:

\begin{equation}\label{eq:powerlaw}
    \Phi_{\gamma}\left(E_{\gamma}\right) = \Phi_{0}\left(\frac{E_{\gamma}}{E_{\mathrm{piv}}}\right)^{-\Gamma},
\end{equation}
where $\Phi_{\gamma}\left(E_{\gamma}\right)$ is the $\gamma$-ray flux at the $\gamma$-ray energy $E_{\gamma}$, $\Phi_{0}$ is the flux normalization at a pivot energy $E_{\mathrm{piv}}$, and $\Gamma$ is the spectral index. $\Phi_{0}$ and $\Gamma$ are set as free parameters to be determined by the likelihood fit, and $E_{\mathrm{piv}}$ is fixed at 20~TeV. {For the pivot energy, we have scanned through the entire energy range from which we have obtained minimum correlation between the free parameters at $20~\mathrm{TeV}$.}

Once the optimal number of point and extended sources is determined from the systematic source search, more realistic spectral models are applied and tested. Power-law with an exponential cutoff and log parabola models have been used:

\begin{eqnarray}
\begin{array}{cc} \label{eq:cutoffpowerlaw}
\Phi_{\gamma}\left(E_{\gamma}\right) = \Phi_{0}\left(\frac{E_{\gamma}}{E_{\mathrm{piv}}}\right)^{-\Gamma}\mathrm{exp}\left(-\frac{E_{\gamma}}{E_{c}}\right)
\end{array},
\\
\begin{array}{cc} \label{eq:logparabola}
\Phi_{\gamma}\left(E_{\gamma}\right) = \Phi_{0}\left(\frac{E_{\gamma}}{E_{\mathrm{piv}}}\right)^{-\Gamma-\beta\ln{\left(E_{\gamma}/E_{\mathrm{piv}}\right)}},\end{array}
\end{eqnarray}
where $E_{c}$ of Equation~\ref{eq:cutoffpowerlaw} is the cutoff energy and $\beta$ of Equation~\ref{eq:logparabola} describes the curvature of the power-law spectrum. 

The radially symmetric Gaussian morphology \citep{HAWCEE} is formulated as

\begin{equation} \label{eq:gaussian}
\frac{dN}{d\Omega}=\left(\frac{180}{\pi}\right)^2\frac{1}{2\pi\sigma^2}\exp{\left(-\frac{\theta^2}{2\sigma^2}\right)}
\end{equation}
where $\theta$ and $\sigma$ are the angular distance and Gaussian width as a measure of the size of the extended source model, respectively. $\sigma$ is set as a free parameter during likelihood fits.

\subsection{Naima Framework} \label{sec:naima}

The Naima software~\footnote{Naima; \url{https://github.com/zblz/naima}} \citep{naima} is a non-thermal $\gamma$-ray modeling framework based on Markov chain Monte Carlo methods. It is used to model and fit the spectra of accelerated particles in hadronic and leptonic scenarios, based on the observed electromagnetic data. 
The Naima software provides numerical models for a hadronic scenario based on neutral pion decays from inelastic proton-proton collisions \citep{PiZero} and a leptonic scenario based on inverse Compton scattering of relativistic electrons with low-energy photons and synchrotron emission from relativistic electrons \citep{InverseCompton,Synchrotron}. We have used the Naima framework to generate particle spectra for the hadronic and leptonic scenarios of HAWC~J1844-034. The results are presented in Sections~\ref{subsec:Scenarios} and \ref{subsec:SNR}. 

\section{Results} \label{sec:Results}

\begin{figure}[ht!]
\gridline{
\fig{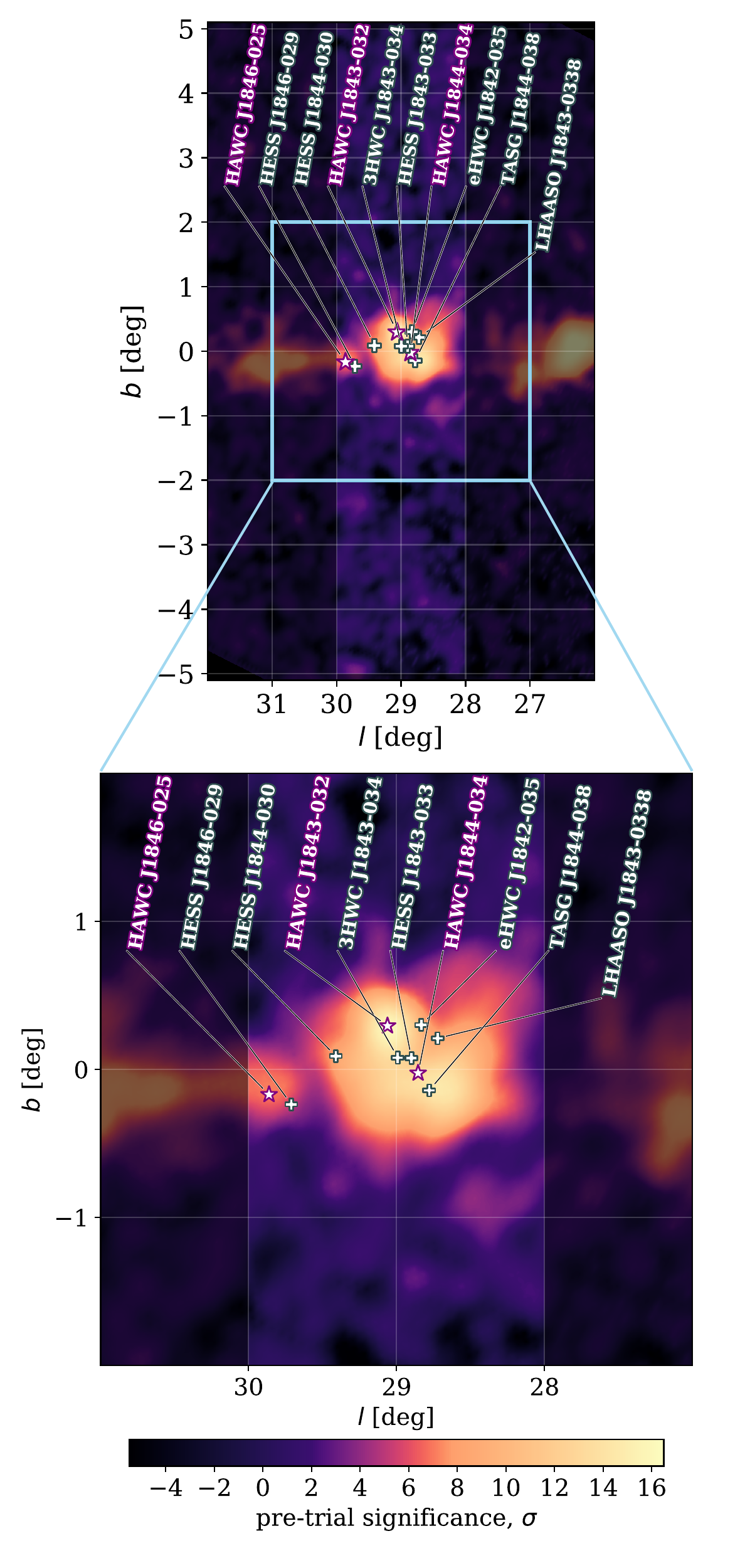}{\columnwidth}{}
}
\caption{1,910 day HAWC significance map in Galactic coordinates highlighting the eHWC~J1842-035 region. The brighter part of the map is the ROI used for the analysis. The labels indicate the known VHE $\gamma$-ray sources within the ROI. Note that 3HWC~J1843-034 indicates the best-fit location of the source found from the HAWC all-energy catalog \citep{3HWC}.
\label{fig:HAWCmap}}
\end{figure}

Figure~\ref{fig:HAWCmap} shows the 1,910~day HAWC significance map of the eHWC~J1842-035 region in Galactic coordinates, where our ROI has been indicated by the brighter rectangular region ($28\degr < l < 30\degr$ and $-5\degr < b < 5\degr$).  
Figure~\ref{fig:HAWCmap} also includes source labels from various $\gamma$-ray observatories such as the High Energy Stereoscopic System (H.E.S.S.), HAWC, Tibet AS-Gamma, and LHAASO \citep{HGPS,eHWC,3HWC,UHELHAASO,Tibet}.

{
\begin{deluxetable*}{@{\hspace{0.3in}}l@{\hspace{0.3in}}|CCCCCCCc}[ht!]
\centering
\tablewidth{0pt}
\tabletypesize{\scriptsize}
\tablenum{1}
\tablecaption{Best-fit parameters found from a multiple source likelihood fit of the nested model in the ROI \label{tab:bestfit}}
\tablehead{
\colhead{Source}  & \multicolumn{1}{|c}{$\alpha$} & 
\colhead{$\delta$} & \colhead{$\Phi_{0}$} & \colhead{$\Gamma$} & \colhead{$\beta$} & \colhead{Size} & \colhead{TS}  & \colhead{{Signif.}} \\
\colhead{} & \multicolumn{1}{|c}{[deg]} & 
\colhead{[deg]} & \colhead{[(TeV cm$^{2}$ s)${}^{-1}$]} & \colhead{} & \colhead{} & \colhead{[deg]} & \colhead{} & \colhead{}
} 
\renewcommand{\arraystretch}{1.4}
\startdata
HAWC~J1844-034 & $281.02^{+0.05}_{-0.05}$ & $-3.64^{+0.05}_{-0.04}$ & $1.32^{+0.10}_{-0.09}{} \times 10^{-14}$ & $2.72^{+0.21}_{-0.22}{}$ & $0.51^{+0.21}_{-0.21}$ & $0.48^{+0.02}_{-0.02}$ & $718.49$ & {$26.26$} \\
HAWC~J1843-032 & $280.83^{+0.04}_{-0.03}$ & $-3.31^{+0.04}_{-0.04}$ & $9.31^{+2.12}_{-1.70} \times 10^{-16}$ & $3.34^{+0.37}_{-0.37}$ & \nodata & \nodata & $33.39$ & {$4.89$} \\
HAWC~J1846-025 & $281.61^{+0.06}_{-0.06}$ & $-2.81^{+0.08}_{-0.08}$ & $4.33^{+1.87}_{-1.29} \times 10^{-16}$ & $2.94^{+0.50}_{-0.48}$ & \nodata & \nodata & $18.67$ & {$3.32$} \\
\enddata
\tablecomments{The listed values are the best-fit parameters and their statistical uncertainties. $\alpha$ and $\delta$ are right ascension and declination, respectively. {Signif. is the post-trial significance.} $\Phi_{0}$ is the flux normalization at a pivot energy of 20~TeV. Size is the 1$\sigma$ radius of the extended Gaussian morphological model used to fit HAWC~J1844-034.}
\end{deluxetable*}}

\begin{figure*}[ht!]
\centering
\gridline{\fig{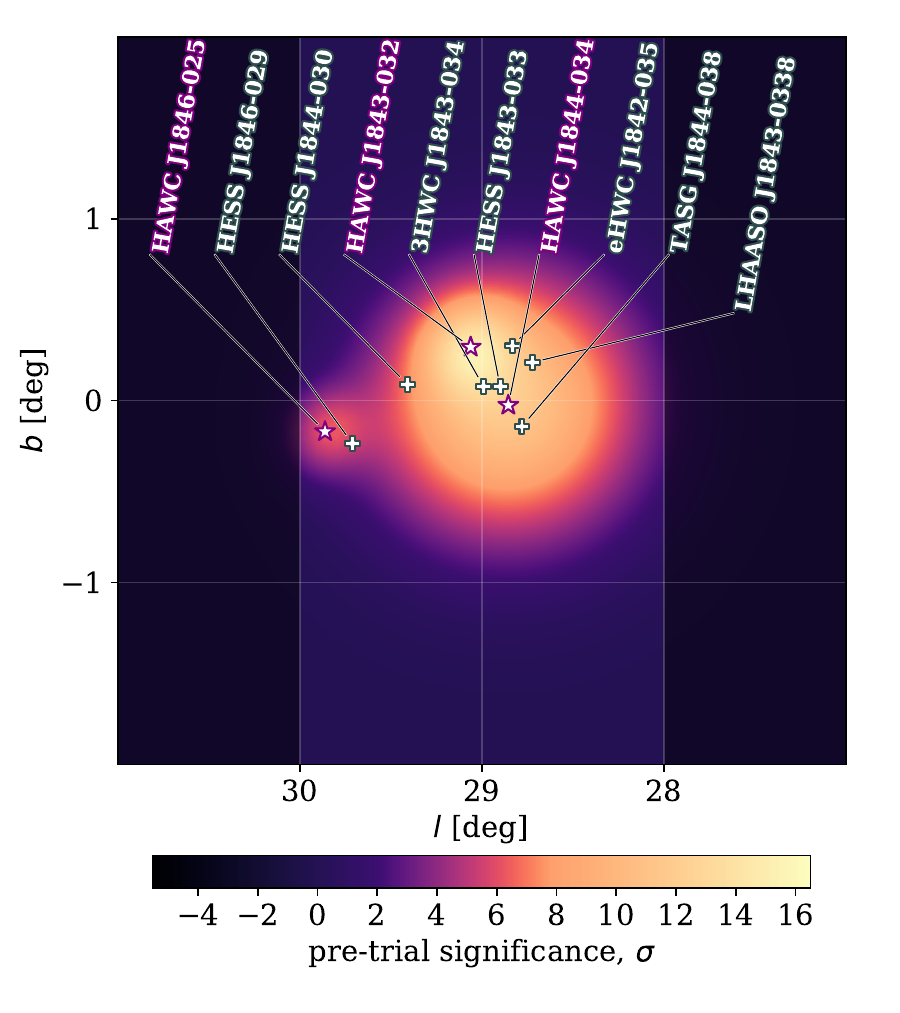}{\columnwidth}{(a)} \fig{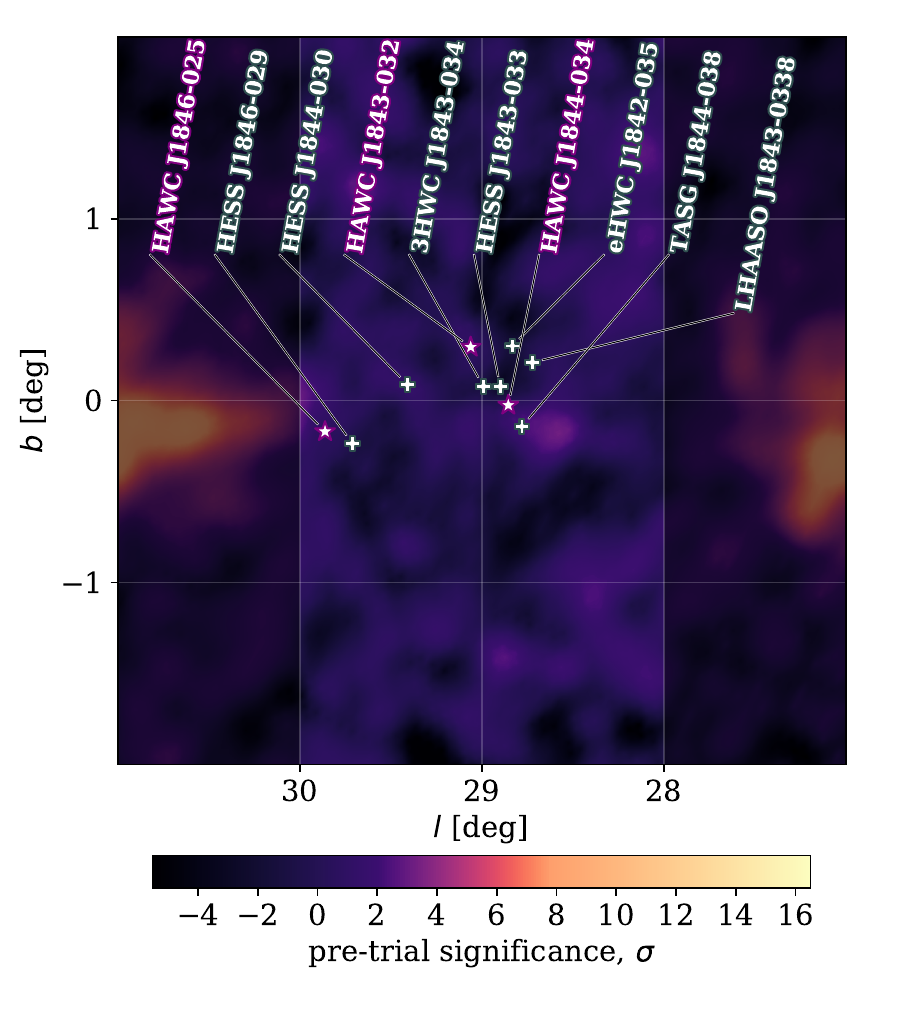}{\columnwidth}{(b)}}
\caption{(a): A significance map of the best-fit model described in Table~\ref{tab:bestfit}. (b): A residual map produced by subtracting the best-fit model. \label{fig:residuals}}
\end{figure*}

From a systematic source search in the ROI, we have found three sources as the statistical optimum, namely, HAWC~J1843-032 and HAWC~J1846-025 as point sources, and HAWC~J1844-034 as an extended source fitted with a radially symmetric Gaussian morphology (Equation~\ref{eq:gaussian}). We have used a power-law spectral model for each of the point sources and a log parabolic spectral model for HAWC~J1844-034. The best-fit parameters at the pivot energy of 20~TeV along with their TS {and post-trial significance values} can be found in Table~\ref{tab:bestfit}. Among the three sources we have found during the systematic source search, HAWC~J1844-034 is by far the brightest and most dominant source with the $1\sigma$ upper energy bound of the observed photons reaching 175~TeV, which is comparable to 260~TeV photons observed by LHAASO \citep{UHELHAASO}. Therefore, we have conducted further analyses on HAWC~J1844-034, which will be discussed in Section~\ref{sec:Discussion}. {Note that the upper energy bound has been obtained by profiling the likelihood of the best-fit spectrum multiplied by a step function while having the bound energy as the only free parameter. The $1\sigma$ level corresponds to the estimate at $\Delta\mathrm{TS}=1$ \citep{J2019}.} {On the other hand, both HAWC~J1843-032 ($4.89\sigma$) and HAWC~J1846-025 ($3.32\sigma$) have their significance $<5\sigma$. Hence, HAWC~J1843-032 and HAWC~J1846-025 are gamma-ray candidates that we have found during the modeling of HAWC~J1844-034 region.}

\begin{figure}[ht]
\fig{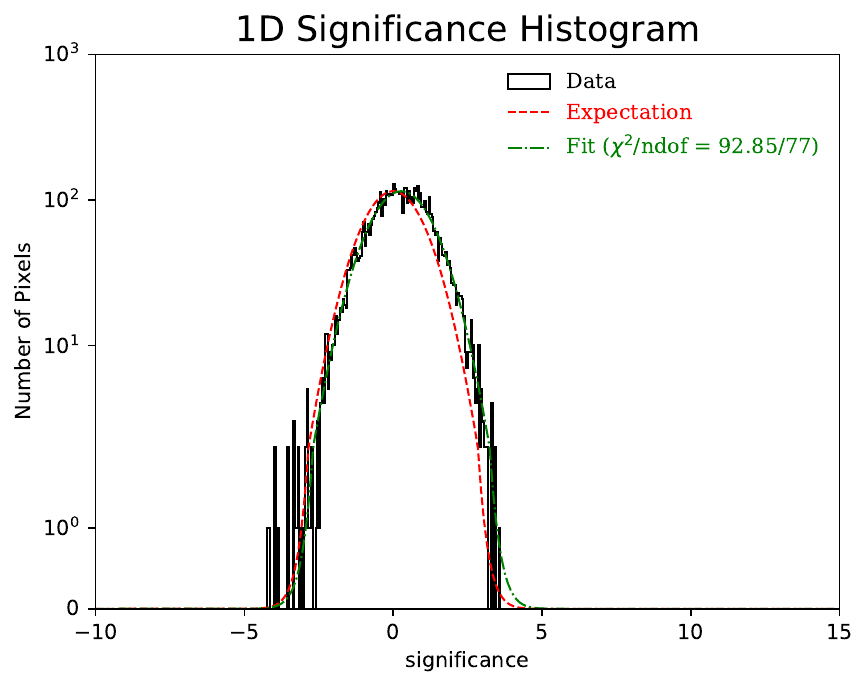}{\columnwidth}{}
\caption{A one dimensional significance distribution histogram collected per pixel of Figure~\ref{fig:residuals}~(b). The expected background-only distribution is described by the red Gaussian curve. {The green line is the best-fit Gaussian curve of the data present in the range between -2 and 2.} \label{fig:sighist}}
\end{figure}

In Figure~\ref{fig:residuals}~(b), a residual map is presented after fitting and subtracting the best-fit model (Figure~\ref{fig:residuals}~(a)) from Figure~\ref{fig:HAWCmap}. 
{Also, Figure~\ref{fig:sighist} is a significance histogram that shows the computed distribution of pixels in the ROI of Figure~\ref{fig:residuals}~(b).}
{The histogram is used to look for obvious signatures of overfitting or underfitting of the $\gamma$-ray excess in the ROI, neither of which are present in the distribution. Also for the histogram, we have computed chi square per number of degrees of freedom of $92.85/77\approx1.20$ for the best-fit Gaussian curve in green.}

\section{Discussions}
\label{sec:Discussion}

\subsection{Spectral Properties of HAWC~J1844-034} \label{subsec:SpectMorph}

\begin{figure*}[ht!]
\fig{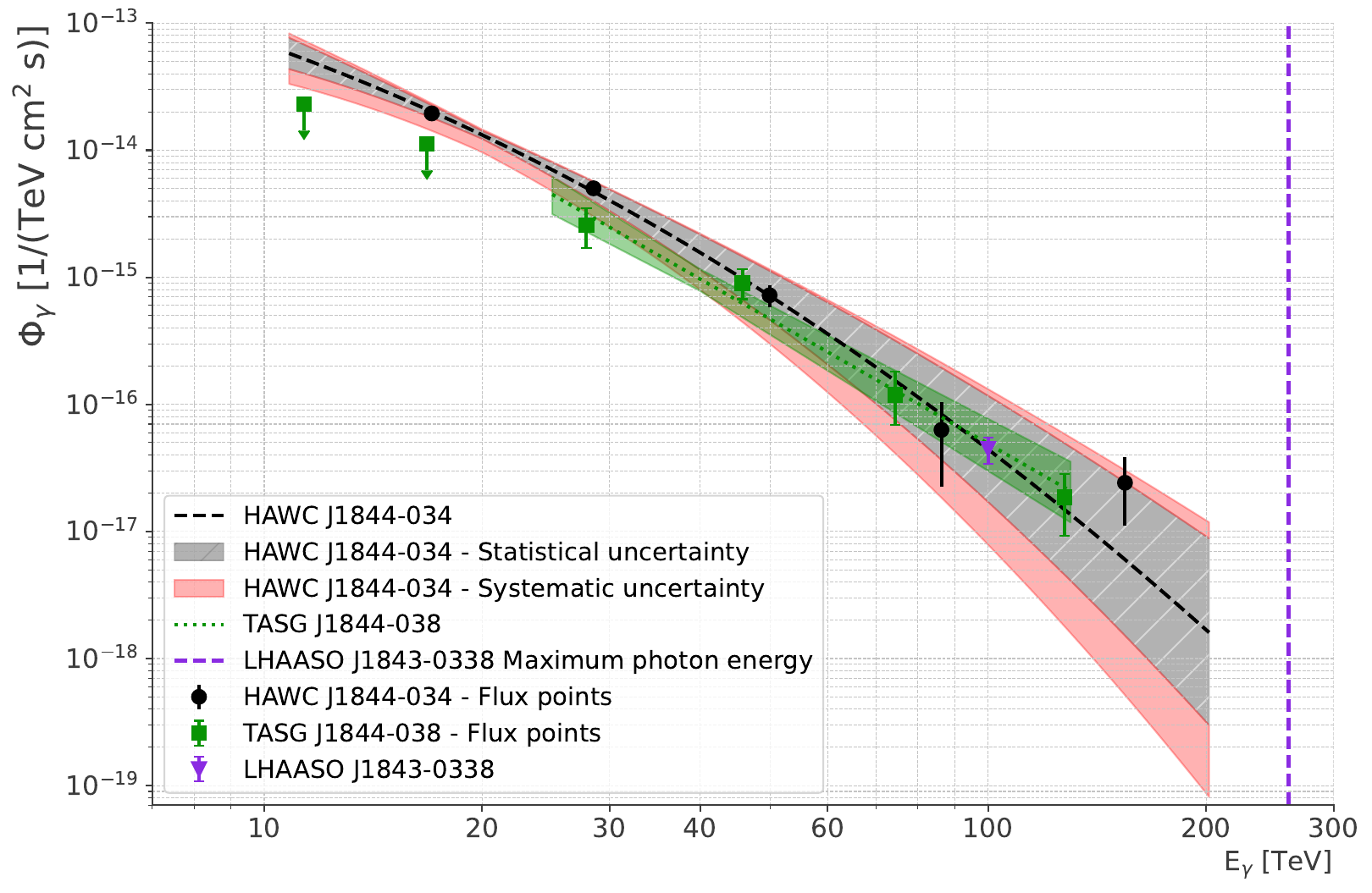}{2.0\columnwidth}{}
\caption{{The best-fit spectrum of HAWC~J1844-034.} \label{fig:Spectrum}}
\end{figure*}

Figure~\ref{fig:Spectrum} shows the spectrum of HAWC~J1844-034. The gray and red bands represent the statistical and systematic uncertainties, respectively. The various systematic uncertainty components have been summed in quadrature (later discussed in Section~\ref{subsec:SystematicUncertainties}). We have calculated the flux points of HAWC~J1844-034 (black) by refitting $\Phi_{0}$ at each energy bin while maintaining the best-fit parameters for the spectral shape, $\Gamma$ and $\beta$ \citep{HAWCEE}.

On the plot, the TeV $\gamma$-ray flux data points from Tibet AS-Gamma (green) and LHAASO (purple) at 100~TeV have also been added for comparison \citep{UHELHAASO,Tibet}.
The fitted HAWC spectrum is in good agreement with the results from both LHAASO and Tibet AS-Gamma. According to the HAWC results, the fall off in the spectrum is gradual, which may be consistent with 260~TeV photons observed by LHAASO (purple dashed line) \citep{UHELHAASO}. The best-fit spectrum of TASG~J1844-038 has been observed with a power-law function in $25~\mathrm{TeV} < E_{\gamma} < 130~\mathrm{TeV}$ \citep{Tibet}. In this energy range, TASG~J1844-038 has four flux points exceeding $2\sigma$, which correspond well with the best-fit spectrum of HAWC~J1844-034. The best-fit power-law spectrum for TASG~J1844-038 has a spectral index of $\Gamma=3.26\pm0.30$, which falls more sharply than the index of HAWC~J1844-034 $\Gamma=2.72^{+0.21}_{-0.22}$ at 20~TeV \citep{Tibet}. This may be due to the different conditions we are performing the fits in such as the energy range, pivot energy, and spectral model. However, there is sufficient overlap despite having different conditions as shown in Figure~\ref{fig:Spectrum}. Also, at the pivot energy of TASG~J1844-038 ($E_{\mathrm{piv}}=40~\mathrm{TeV}$), the spectral index of HAWC~J1844-034 is found to be $\Gamma=3.07^{+0.36}_{-0.37}$, which is consistent with the results from Tibet AS-Gamma. Furthermore, the sources found by the three gamma-ray observatories are spatially consistent given their angular resolutions: $\sim0\fdg1 - 0\fdg6$ for $\geq10~\mathrm{TeV}$ (HAWC), $0\fdg24 - 0\fdg3$ for $100~\mathrm{TeV}$ (LHAASO), and $\sim0\fdg28$ for $100~\mathrm{TeV}$ (Tibet AS-Gamma) \citep{HAWCEE, LHAASOAngularResolution, Tibet}.

\subsection{Extended Emission from HAWC~J1844-034}

\begin{figure*}[ht]
\fig{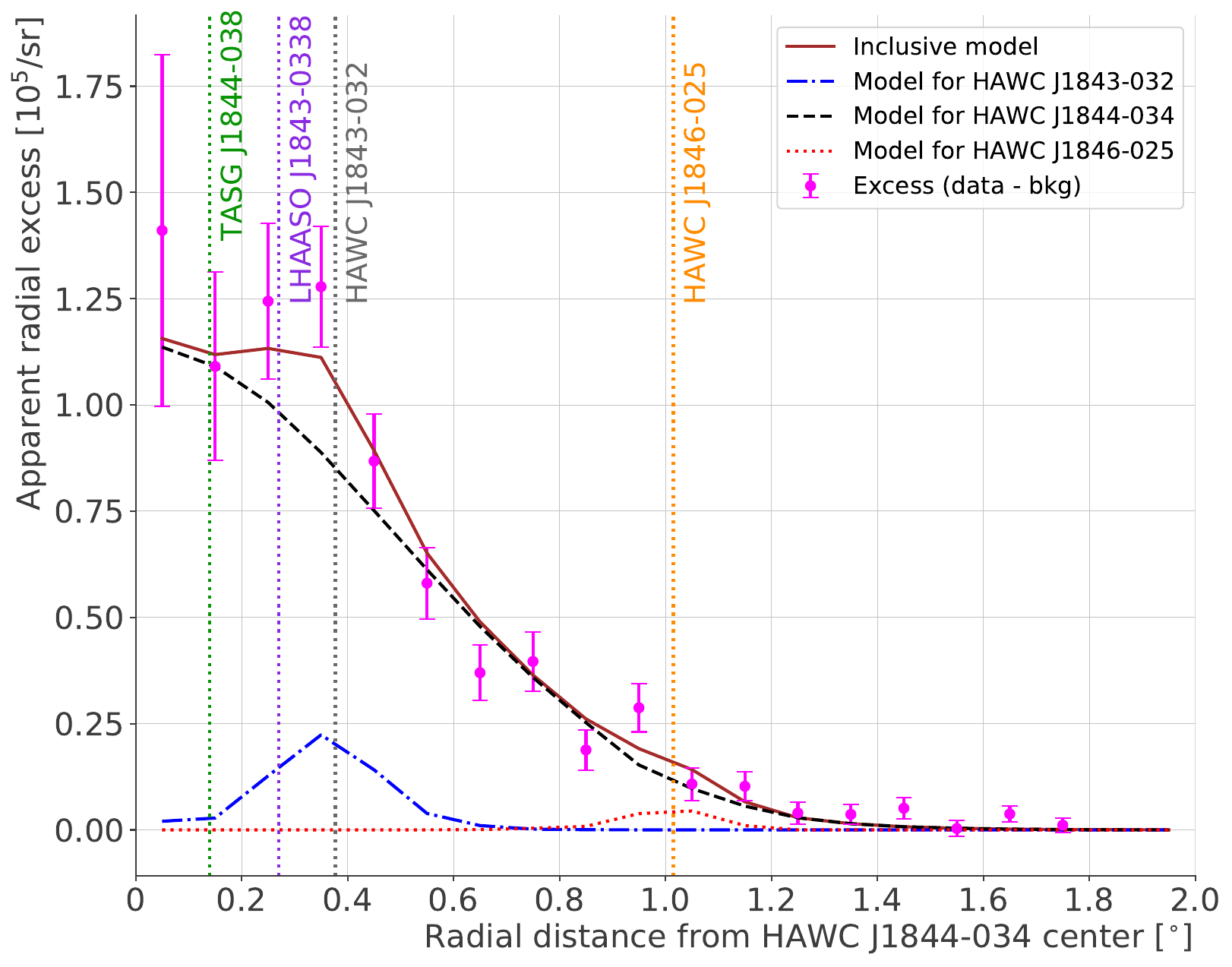}{1.7\columnwidth}{}
\caption{The radial profile from the best-fit location of HAWC~J1844-034 as center. The dotted lines are radial distances of the best-fit positions of the sources from the center of HAWC~J1844-034.
\label{fig:radial_profile}}
\end{figure*}

Figure~\ref{fig:radial_profile} displays the radial distribution of $\gamma$-ray excess profile from the best-fit location of HAWC~J1844-034 (magenta points). The brown solid line represents the overall best-fit model that includes all three HAWC sources of the region. The black dashed, blue dot-dashed, and red dotted lines represent the best-fit models for each source, and the dotted vertical lines represent the distances of each labeled source from HAWC~J1844-034. Note that TASG~J1844-038 (green) and LHAASO~J1843-0338 (purple) are not included in the inclusive model. 
The inclusive model shows a good match with the excess. Also, TASG~J1844-038 presents a good spatial agreement with HAWC~J1844-034. Conversely, LHAASO~J1843-0338 is positioned between the peak of HAWC~J1844-034 and HAWC~J1843-032, which corresponds to a plateau of the overall excess peak. 

\subsection{Systematic Uncertainties} \label{subsec:SystematicUncertainties}

{
\begin{deluxetable*}{l@{\hspace{1.0in}}|CCCC|CC|CC}[ht!]
\tablewidth{0pt}
\tabletypesize{\scriptsize}
\tablenum{2}
\tablecaption{Systematic uncertainties \label{tab:systematics}}
\tablehead{
\colhead{} & \multicolumn{4}{|c}{HAWC~J1844-034} & \multicolumn{2}{|c}{HAWC~J1843-032} & \multicolumn{2}{|c}{HAWC~J1846-025} \\
\colhead{Uncertainty origin} & \multicolumn{1}{|c}{$\Phi_{0}$} & \colhead{$\Gamma$} & 
\colhead{$\beta$} & \colhead{Size} & \multicolumn{1}{|c}{$\Phi_{0}$} & \colhead{$\Gamma$} & \multicolumn{1}{|c}{$\Phi_{0}$} & \colhead{$\Gamma$} \\
\colhead{} & \multicolumn{1}{|c}{[(TeV cm$^{2}$ s)${}^{-1}$]} & \colhead{} & 
\colhead{} & \colhead{[deg]} & \multicolumn{1}{|c}{[(TeV cm$^{2}$ s)${}^{-1}$]} & \colhead{} & \multicolumn{1}{|c}{[(TeV cm$^{2}$ s)${}^{-1}$]} & \colhead{} 
} 
\renewcommand{\arraystretch}{1.4}
\startdata
Late light & ${}^{+1.08}_{-2.21} \times 10^{-15}$ & ${}^{+0.22}_{-0.08}$ & ${}^{+0.10}_{-0.27}$ & ${}^{+0.00}_{-0.00}$ & ${}^{+3.87}_{-6.73} \times 10^{-17}$ & ${}^{+0.01}_{-0.02}$ & ${}^{+3.75}_{-7.01} \times 10^{-17}$ & ${}^{+0.08}_{-0.03}$ \\
Charge uncertainty & $-7.48 \times 10^{-16}$ & $+0.08$ & $-0.04$ & $+0.00$ & $-4.75 \times 10^{-17}$ & $+0.03$ & $-2.26 \times 10^{-17}$ & $+0.03$ \\
PMT efficiency & $+4.67 \times 10^{-16}$ & ${}^{+0.02}_{-0.04}$ & $-0.04$ & ${}^{+0.00}_{-0.00}$ & $+6.13 \times 10^{-17}$ & ${}^{+0.01}_{-0.02}$ & $+2.67 \times 10^{-17}$ & ${}^{+0.02}_{-0.03}$ \\
PMT threshold & ${}^{+1.36}_{-1.11} \times 10^{-16}$ & ${}^{+0.09}_{-0.05}$ & ${}^{+0.06}_{-0.07}$ & ${}^{+0.00}_{-0.00}$ & ${}^{+1.86}_{-1.00} \times 10^{-18}$ & ${}^{+0.04}_{-0.02}$ & $+9.06 \times 10^{-18}$ & $+0.03$ \\
{DBE} & ${-2.70} \times 10^{-15}$ & ${-0.11}$ & $+0.37$ & ${-0.08}$ & ${-1.52} \times 10^{-17}$ & ${-0.02}$ & ${-8.06 \times 10^{-19}}$ & ${+0.06}$ \\
{ROI} & ${+7.17 \times 10^{-16}}$ & ${+0.00}$ & ${-0.02}$ & ${+0.03}$ & ${-2.78 \times 10^{-18}}$ & ${-0.02}$ & ${+1.61 \times 10^{-16}}$ & ${-0.34}$ \\
\hline
Quadratic summed & ${}^{{+1.38}}_{{-3.57}} \times 10^{-15}$ & ${}^{+0.25}_{{-0.15}}$ & ${}^{+0.39}_{{-0.29}}$ & ${}^{+0.03}_{{-0.08}}$ & ${}^{+7.25}_{{-8.38}} \times 10^{-17}$ & ${}^{+0.05}_{{-0.04}}$ & ${}^{{+1.68}}_{{-0.74}} \times 10^{-16}$ & ${}^{{+0.11}}_{-0.34}$ \\ 
\enddata
\end{deluxetable*}}

Factors that may contribute towards the systematics are listed in Table~\ref{tab:systematics}. The first four components in the table arise from various detector and simulation effects \citep[see][]{HAWCCrab, HAWCEE}. 
The DBE is a systematic effect arising from any unresolved sources and the true Galactic diffuse emission. We have calculated this effect by eadding a Gaussian background model, along the Galactic equator {with the fixed width of $1\degr$ \citep{J1825, HAWCGDE}}, in the nested fit. We have compared the best-fit parameters of the three sources and added the differences as a systematic component. {A power-law spectral model (Equation~\ref{eq:powerlaw}) with an index of $2.75$ is assumed for the DBE based on \citet{GDEMilagro, J1908}.} 

{Finally, our choice of ROI has been considered as a systematic uncertainty. Our ROI has been chosen as the best compromise between considering a large enough region to include the extended gamma-ray excess of HAWC~J1844-034 and minimizing the number of sources we would have to simultaneously fit. During the analysis, the ROI is predetermined before the systematic source search, hence we have used a larger ROI of $28\degr < l < 30.3\degr$ to see how much effect this would have on the source of our main interest, HAWC~J1844-034. Although extending the ROI has shown a notable effect on HAWC~J1846-025, there is a minimal effect on HAWC~J1844-034.}

\subsection{Leptonic versus hadronic scenarios} \label{subsec:Scenarios}

The nature of the gamma-ray emission from HAWC~J1844-034 has been modeled and studied to test if this source is a potential hadronic accelerator. In the hadronic scenario, the emission of $\gamma$-rays is dominated by the process $\pi^{0} \rightarrow 2\gamma$. If the neutral pions are produced from the collisions of ambient protons that are accelerated by HAWC~J1844-034 \citep{PiZero}, then the observed gamma-ray spectrum can be used to constrain the proton spectrum of HAWC~J1844-034 \citep{PeVatronSignature}. The hadronic scenario is powered by the proton-proton scatterings of accelerated protons and protons of the ambient gas. {During the fit, we have assumed the proton density of 1~cm$^{-3}$.}

In the case of the leptonic scenario, the TeV $\gamma$-ray emission is produced mostly from the inverse Compton scattering between low-energy ambient photons and high-energy electrons accelerated by HAWC~J1844-034 \citep{InverseCompton}. Due to the Klein-Nishina effects, the $\gamma$-ray spectrum originating from the inverse Compton scattering has a spectral cutoff at tens of TeV \citep{J2032}. For the inverse Compton scattering of the leptonic scenario, we used cosmic microwave background (CMB) photons as seed photons that are up-scattered by the accelerated electrons. 

\begin{figure}[ht!]
\fig{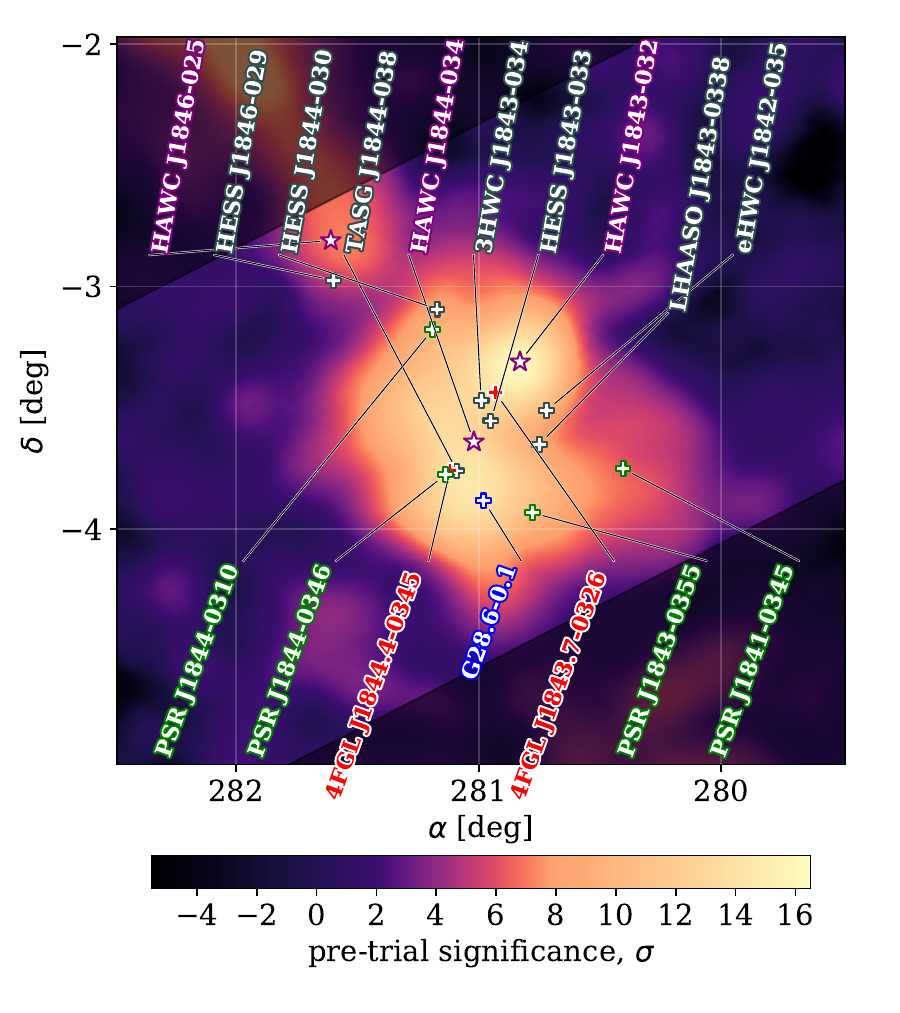}{\columnwidth}{}
\caption{A significance map of the HAWC~J1844-034 region with TeV gamma-ray sources labeled at the top \citep{HGPS,Tibet,eHWC,3HWC,UHELHAASO}, Fermi-LAT sources in red \citep{4FGL}, and known pulsars and SNR in green and blue, respectively \citep{ATNF,ASCA,Chandra2}. \label{fig:CounterpartsMap}}
\end{figure}

Figure~\ref{fig:CounterpartsMap} presents the locations of the sources inside our ROI. The labels at the top of the figure are the VHE $\gamma$-ray sources found by H.E.S.S., Tibet AS-Gamma, LHAASO, and previously detected sources from HAWC catalogs \citep{HGPS,Tibet,UHELHAASO}. The purple labels indicate the sources specifically found by this work. The lower red labels indicate GeV $\gamma$-ray sources observed by Fermi-LAT \citep[denoted as 4FGL]{4FGL}, the green labels are for pulsars from the ATNF pulsar catalog \citep{ATNF}, and the blue label marks a known supernova remnant SNR~G28.6-0.1 \citep{NVSS,ASCA,Chandra}. 

{
\begin{deluxetable*}{c|CCCC}[ht!]
\tablewidth{0pt}
\tabletypesize{\scriptsize}
\tablenum{3}
\tablecaption{A list of known $\gamma$-ray sources near HAWC~J1844-034 \label{tab:GammaSources}}
\tablehead{
\colhead{Source} & \multicolumn{1}{|c}{$\alpha$ [deg]} & \colhead{$\delta$ [deg]} & \colhead{$\Delta$ [deg]} & \colhead{Size [deg]}
} 
\renewcommand{\arraystretch}{1.4}
\startdata
HAWC~J1844-034 & $281.02{^{+0.05}_{-0.05}}$ & $-3.64{^{+0.05}_{-0.04}}$ & \nodata & {$0.48^{+0.02}_{-0.02}$} \\
\hline
LHAASO~J1843-0338 & $280.75$ & ${-3.65}$ & $0.27$ & $0.3$ \\
TASG~J1844-038 & $281.09{^{+0.10}_{-0.10}}$ & $-3.76{^{+0.09}_{-0.09}}$ & $0.14$ & {$0.34\pm0.12$} \\
HESS~J1843-033 & $280.95$ & $-3.55$ & $0.11$ & {$0.24\pm0.06$} \\
HESS~J1844-030 & $281.17$ & $-3.10$ & $0.56$ & {$0.02\pm0.013$} \\
4FGL~J1843.7-0326 & $280.93$ & $-3.44$ & $0.22$ & \nodata \\
4FGL~J1844.4-0345 & $281.12$ & $-3.75$ & $0.15$ & \nodata \\
\enddata
\tablecomments{$\Delta$ is the angular distance from the center of HAWC~J1844-034. The size of LHAASO~J1843-0338 is referred from the extension template of $0\fdg3$ \citep{UHELHAASO}. The size of TASG~J1844-038 does not include the PSF contribution \citep[see Equation~1]{Tibet}. HESS~J1843-033 consists of two Gaussian peaks \citep{HGPS}. {The positional error of LHAASO J1843-0338 was not reported.}}
\end{deluxetable*}}

\begin{figure*}[ht!]
\fig{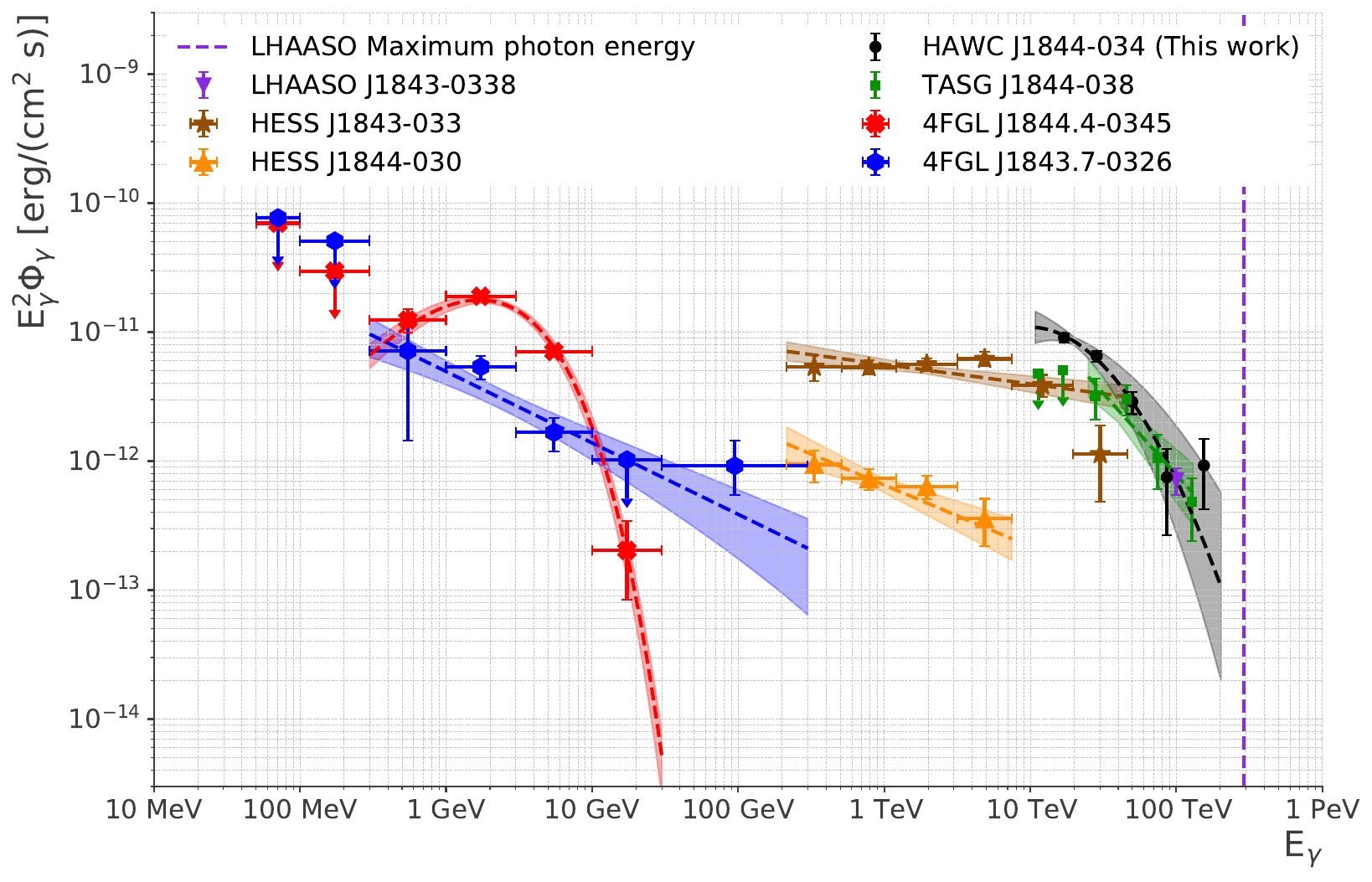}{2.0\columnwidth}{}
\caption{A collection of spectra for $\gamma$-ray sources within the vicinity of HAWC~J1844-034. The results have been collected from Fermi-LAT \citep{4FGL}, H.E.S.S. \citep{HGPS}, Tibet AS-Gamma \citep{Tibet}, LHAASO \citep{UHELHAASO}, and HAWC.\label{fig:Counterparts}}
\end{figure*}

Figure~\ref{fig:Counterparts} shows the spectra of the GeV to TeV $\gamma$-ray sources within the extension of HAWC~J1844-034 as shown in Figure~\ref{fig:CounterpartsMap}. HESS~J1843-033 has been considered as the counterpart of TASG~J1844-038 \citep{Tibet}. The observed spectrum of HESS~J1843-033 (brown) has a similar amplitude as that of HAWC~J1844-034 (black) and TASG~J1844-038 (green), so it has also been assumed as the sub-TeV counterpart in our analysis. The flux points of HESS~J1844-030 (orange) and 4FGL~J1843.7-0326 (blue) are smoothly connected, so it is likely that these two observations come from the same source. However, it is difficult to find an association of these two sources with HAWC~J1844-034 (as well as TASG~J1844-038 and LHAASO~J1843-0338) because HESS~J1844-030 has not been observed with significant emission at energies above 10~TeV \citep{HGPS}. The spectrum of 4FGL~J1844.4-0345 (red) has been found to follow a log parabolic shape in the GeV range and cuts off sharply, not detected above 100~GeV from Fermi-LAT \citep{4FGL}.
Therefore, it is difficult to identify a source in the GeV energies that shows clear association with HAWC~J1844-034 at this time.

The two scenarios have been fitted with the Naima framework using only the TeV gamma-ray flux data from HESS~J1843-033 \citep{HGPS}, TASG~J1844-038 \citep{Tibet}, LHAASO~J1843-0338 \citep{UHELHAASO}, and HAWC~J1844-034 (this work). 
The particle spectra of each scenario have been assumed to be a power law with an exponential cutoff with a pivot energy of 7~TeV. 
\begin{eqnarray}\label{eq:EnergyDist}
    \rho_{\mathrm{P}}(E_{\mathrm{P}}) = \rho_{0,\mathrm{P}\left(1\mathrm{kpc}\right)}\left(\frac{E_{\mathrm{P}}}{7~\mathrm{TeV}}\right)^{-\Gamma_{\mathrm{P}}} \nonumber \\
    \times \mathrm{exp}\left(-\frac{E_{\mathrm{P}}}{E_{c,\mathrm{P}}}\right)
\end{eqnarray}
where $\rho_{\mathrm{P}}(E_{\mathrm{P}})$ is the spectrum of the accelerated particles, either electrons or protons, $E_{\mathrm{P}}$ is energy of accelerated particles, $\rho_{0,\mathrm{P}\left(1\mathrm{kpc}\right)}$ is the value of $\rho_{\mathrm{P}}(E_{\mathrm{P}})$ at the distance of 1~kpc and the pivot energy of 7~TeV.

\begin{deluxetable}{@{\hspace{0.25in}}c@{\hspace{0.25in}}|C|C}[ht]
\tablewidth{0pt}
\tabletypesize{\scriptsize}
\tablenum{4}
\tablecaption{Best-fit parameters for the hadronic and leptonic scenarios \label{tab:scenarios}}
\tablehead{
\colhead{} & \multicolumn{1}{|c}{Hadronic Scenario} & \multicolumn{1}{|c}{Leptonic Scenario}
} 
\renewcommand{\arraystretch}{1.4}
\startdata
$\rho_{0,\mathrm{P}\left(1\mathrm{kpc}\right)}$ [1/eV] & {$1.85^{+0.34}_{-0.37}\times10^{34}$} & $7.04^{+1.17}_{-1.13}\times10^{31}$ \\
$\Gamma_{\mathrm{P}}$ & $1.56^{+0.11}_{-0.17}$ & $2.17^{+0.14}_{-0.20}$ \\
$E_{c,\mathrm{P}}$ [TeV] & $169^{+32}_{-35}$ & $72^{+13}_{-13}$ \\
{BIC} & {$85.9$} & {$84.4$} \\
\enddata
\tablecomments{$\rho_{0,\mathrm{P}\left(1\mathrm{kpc}\right)}$ is normalization at a source distance of 1~kpc, $\Gamma_{\mathrm{P}}$ is spectral index, and $E_{c,\mathrm{P}}$ is cutoff energy.}
\end{deluxetable}

Table~\ref{tab:scenarios} shows the best-fit values for the proton spectrum of the hadronic scenario and the electron spectrum of the leptonic scenario. The three free parameters of Equation~\ref{eq:EnergyDist} are listed for each scenario in the table. {The obtained spectral index of 1.56 for the hadronic scenario may seem hard considering other hadronic spectra in general. However, hadronic scenarios with such extreme spectral indices are discussed in some cases \citep{HESSJ1702}.} The best-fit cutoff energy for the hadronic scenario at 169~TeV is too low for HAWC~J1844-034 to be defined as a PeVatron. 
Note that the particle spectra have been fitted while assuming HAWC~J1844-034 is located at a unit distance of 1 kpc since the exact counterpart source has not been identified. The source distance is later accounted for with a known source distance in Section~\ref{sec:GammaEmissionModeling} when we try to identify a counterpart source.

\begin{figure*}[ht!]
\fig{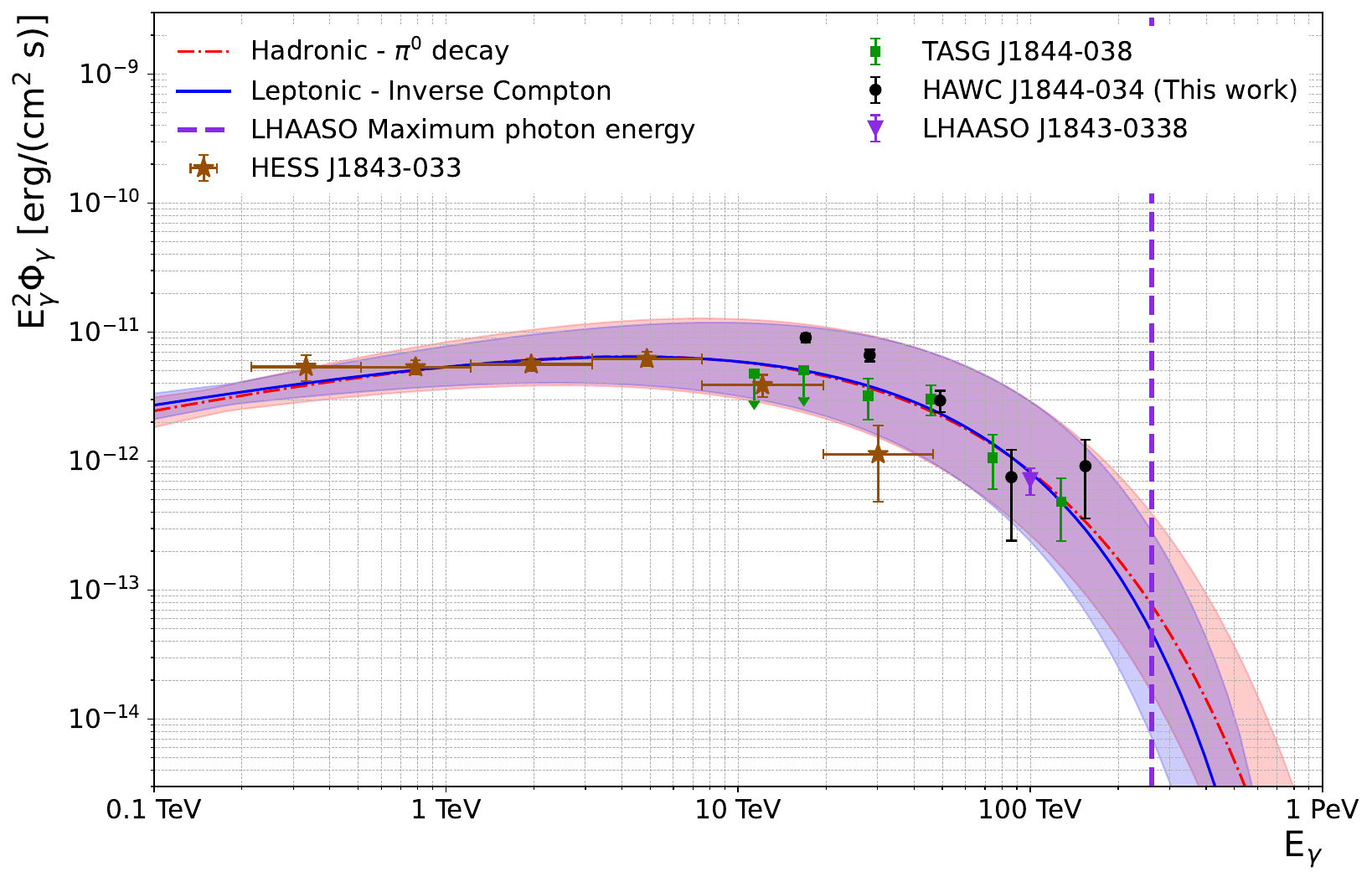}{2.0\columnwidth}{}
\caption{A spectral distributions plot of the best-fit leptonic (blue) and hadronic (red) scenarios that have been fitted using the TeV gamma-ray flux points speculated of having the same original source \citep{HGPS,Tibet,UHELHAASO}. The error bands indicate statistical errors. \label{fig:TeVScenario}}
\end{figure*}

In Figure~\ref{fig:TeVScenario}, we have the data points from LHAASO~J1843-0338 (purple), HESS~J1843-033 (brown), TASG~J1844-038 (green) and HAWC~J1844-034 (black). The dashed purple vertical line is the maximum photon energy as observed by LHAASO on this source. The $\gamma$-ray spectra obtained from fitting the data points with the Naima framework are presented as blue and red lines for the leptonic and hadronic scenarios, respectively, with their statistical uncertainties shown as lighter bands. The overall trend of the spectra are very similar at energies below 100~TeV. At higher energies, the two spectra start to diverge as the leptonic spectrum cuts off at lower energies. {From the Naima fit, the leptonic model is slightly more favorable than the hadronic model with a difference in BIC of 1.5.}

{
\begin{deluxetable*}{c|CCCC|CC}[ht!]
\tablewidth{0pt}
\tabletypesize{\scriptsize}
\tablenum{5}
\tablecaption{{Various properties of four pulsars found near HAWC~J1844-034 used for calculations \citep{ATNF,PSRJ1844Distance}}
\label{tab:PulsarsCharacteristics}}
\tablehead{
\colhead{} & \multicolumn{2}{|c}{} & \colhead{Spin-Down} & \colhead{} & \multicolumn{1}{|c}{Angular} & \colhead{{$\gamma$-ray}} \\
\colhead{Pulsar} & \multicolumn{1}{|c}{Right Ascension} & \colhead{Declination} & \colhead{Power} & \colhead{Distance} & \multicolumn{1}{|c}{Distance} & \colhead{{Luminosity}} \\
\colhead{} & \multicolumn{1}{|c}{$\alpha$ [deg]} & \colhead{$\delta$ [deg]} & \colhead{$\dot{E}$ [erg/s]} & \colhead{$d_{\mathrm{src}}$ [kpc]} & \multicolumn{1}{|c}{$\Delta$ [deg]} & \colhead{{$L_{\gamma}^{\mathrm{HAWC}}$ [erg/s]}}
} 
\renewcommand{\arraystretch}{1.4}
\startdata
PSR~J1841-0345 & 280.4112 & -3.8118 & $2.7 \times ~ 10^{35}$ & $3.776$ & $0.63$ & {$2.14^{+0.44}_{-0.27} \times ~ 10^{34}$} \\
PSR~J1844-0346 & 281.13704 & -3.7752 & $4.2 \times ~ 10^{36}$ & $4.3$ & $0.18$ & {$2.78^{+0.53}_{-0.35} \times ~ 10^{34}$} \\
PSR~J1843-0355 & 280.77776 & -3.93239 &  $1.77 \times ~ 10^{34}$ & $5.798$ & $0.38$ & {$5.05^{+0.97}_{-0.64} \times ~ 10^{34}$} \\
PSR~J1844-0310 & 281.1895 & -3.1770 & $2.79 \times ~ 10^{33}$ & $5.973$ & $0.49$ & {$5.36^{+1.03}_{-0.67} \times ~ 10^{34}$} \\
\enddata
\tablecomments{Angular distance is the distance from the center of HAWC~J1844-034.}
\end{deluxetable*}}

\section{Origin of the \texorpdfstring{$\gamma$}{g}-ray Emission of HAWC~J1844-034} \label{sec:GammaEmissionModeling}

In this section, we analyze nearby pulsars and a supernova remnant to gauge their potentials as a counterpart to HAWC~J1844-034.

\subsection{Pulsars} \label{subsec:Pulsars}

There are three pulsars that are within the best-fit size of HAWC~J1844-034: PSR~J1843-0355, PSR~J1844-0310, and PSR~J1844-0346 \citep{ATNF}. Also, PSR~J1841-0345 is suggested as a possible counterpart source to LHAASO~J1843-0338 in \citet{LHAASOPSR}. 
Therefore, we have examined these four pulsars as a candidate counterpart source to HAWC~J1844-034.
The properties of the pulsars used in the analysis have been referenced from the ATNF catalog \citep{ATNF} and listed in Table~\ref{tab:PulsarsCharacteristics}. 

In Table~\ref{tab:PulsarsCharacteristics}, we have computed a number of parameters on the four pulsars for comparison. $L_{\gamma}^{\mathrm{HAWC}}$ is the $\gamma$-ray luminosity estimated from integrating the spectrum of HAWC~J1844-034 in Figure~\ref{fig:Spectrum} as shown below:

\begin{equation}
    L_{\gamma}^{\mathrm{HAWC}} = 4\pi d_{\mathrm{src}}^2 \int_{~11~\mathrm{TeV}}^{202~\mathrm{TeV}} dE_{\gamma} ~ E_{\gamma} \Phi_{\gamma}.
\label{eq:luminosity}
\end{equation}

The angular distance is the distance to each pulsar from the best-fit location of HAWC~J1844-034. 

{Assuming the spin-down of a pulsar powers the acceleration of the particles that then produce TeV $\gamma$-rays, the $\gamma$-ray luminosity is required to be less than the spin-down luminosity of a pulsar to be a possible counterpart source \citep{TeVPulsar, J1928, LHAASOPSR}.} {PSR~J1843-0355 and PSR~J1844-0310 both have their spin-down powers that are lower than their $\gamma$-ray luminosities if they were associated with HAWC~J1844-034. Therefore, they are disfavored to be the counterpart source of HAWC J1844-034.}

{PSR~J1841-0345 is located $0\fdg63$ away from HAWC~J1844-034, which is outside of the extension of HAWC~J1844-034. Therefore, we rule out PSR~J1841-0345 from the counterpart source candidates.}

{Finally, PSR~J1844-0346 has its spin-down power that is approximately two orders of magnitude higher than the gamma-ray luminosity. Also, it is located closest to the best-fit positions of HAWC~J1844-034, HESS~J1843-033, and TASG~J1844-038 with the angular distances of $0\fdg18$, $0\fdg28$, and $0\fdg05$, respectively. The angular resolutions for HAWC, H.E.S.S., and Tibet AS-Gamma are $0\fdg1 - 0\fdg6$ for $\geq10~\mathrm{TeV}$, $\sim0\fdg08$ for $\geq0.2~\mathrm{TeV}$, and $\sim0\fdg28$ for $100~\mathrm{TeV}$, respectively \citep{HAWCEE, HGPS, Tibet}. Here, the angular resolution of H.E.S.S. is smaller than the angular distance between HESS~J1843-033 and PSR~J1844-0346. However, HESS~J1843-033 is an extended source with the best-fit size of $0\fdg24$ (Table~\ref{tab:GammaSources}). Furthermore, although the angular distance of $0\fdg18$ is larger than the positional uncertainties given in Table~\ref{tab:GammaSources}, the $0\fdg18$ separation is contained within the $0\fdg48$ extension of HAWC~J1844-034. Therefore, of the four pulsars found near the region of HAWC~J1844-034, PSR~J1844-0346 is considered to be the most favorable candidate according to this work.}

\subsection{SNR~G28.6-0.1} \label{subsec:SNR}

SNR~G28.6-0.1 is a shell-type supernova remnant \citep{ASCA} that is another candidate counterpart source to HAWC~J1844-034. It is located at an angular distance of $0\fdg28$ away from HAWC~J1844-034, which is within the best-fit size of HAWC~J1844-034. SNR~G28.6-0.1 was suggested to be a potential candidate counterpart source for HESS~J1843-033 with VHE gamma-rays originating from leptonic processes due to the observed synchrotron emission \citep{PSRJ1844Distance,ASCA}. Assuming SNR~G28.6-0.1 is responsible for the observed VHE $\gamma$-rays, by using the distance to SNR~G28.6-0.1 of $9.6\pm0.3$~kpc \citep{SNRAge}, the $\gamma$-ray luminosity and the proton density can be estimated as presented in Table~\ref{tab:SNREstimated}. 

{We have calculated the energy budgets corresponding to each scenario at the known distance of SNR~G28.6-0.1 to compare between leptonic and haronic scenarios for the SNR. The proton density in the ambient gas of HAWC~J1844-034, $n_{h,d_{\mathrm{src}}}$, is required for the calculation. We use $N\left(\mathrm{HI}\right) = 1.8 \times 10^{21} ~\mathrm{cm}^{-2}$ as the HI column density and $N\left(\mathrm{H}_{2}\right) = 6.7 \times 10^{21} ~\mathrm{cm}^{-2}$ as the ${\mathrm{H}}_{2}$ column density, collected from the HI4PI survey and FUGIN data, respectively, at the center of HAWC~J1844-034 \citep{HI4PI, FUGIN}. The integration range of the radial velocity that corresponds to the location of SNR~G28.6-0.1 that has been used to calculate the column densities is found to be $77-88~\mathrm{km/s}$ based on \citet{KinDist}.} Assuming the density is spherically uniform within the best-fit size of HAWC~J1844-034, the proton density of the ambient gas can be estimated as
\begin{eqnarray}
    n_{h,d_{\mathrm{src}}} = \frac{\Theta}{V} = \frac{3(N\left(\mathrm{HI}\right)+2N\left(\mathrm{H}_{2}\right))}{4} \nonumber \\ 
    \times \frac{180\degr}{0\fdg48~\pi d_{\mathrm{src}}},
\label{eq:hadron_density}
\end{eqnarray}
where $\Theta$ is the total number of protons, $V$ is the spherical volume of the emission region, and $0\fdg48\pi/180\degr$ is the best-fit size of HAWC~J1844-034 in radians. {The estimated proton density for SNR~G28.6-0.1 is $\sim46~\mathrm{cm^{-3}}$.}

The energy budget for each scenario is computed by integrating the particle spectrum $\rho_{\mathrm{P}}(E_{\mathrm{P}})$ {provided in Table~\ref{tab:scenarios}},

\begin{equation}
    W_{\mathrm{P}} = \int_{~E_{\mathrm{min},\mathrm{P}}}^{E_{\mathrm{max},\mathrm{P}}} dE_{\mathrm{P}}~ E_{\mathrm{P}}\rho_{\mathrm{P}}(E_{\mathrm{P}}),
\label{eq:energybudget}
\end{equation}
where $E_{\mathrm{min},\mathrm{P}}$ and $E_{\mathrm{max},\mathrm{P}}$ are the minimum and maximum energies of the particle spectrum, respectively.

As for the hadronic scenario, we have used 1.22~GeV for the minimum proton energy $E_{\mathrm{min},p}$ since the inclusive $\pi^{0}$ cross section of a proton-proton collision starts to become meaningful at energies above 1.22~GeV \citep[see][Section 2.B.2]{PiZero}. For the maximum proton energy $E_{\mathrm{max},p}$, we have used 2.6~PeV, which is ten times the maximum photon energy observed by LHAASO \citep{UHELHAASO}. 

{
\begin{deluxetable*}{c|CC|CCCCC}[ht!]
\tablewidth{0pt}
\tabletypesize{\scriptsize}
\tablenum{6}
\tablecaption{Computed values for SNR~G28.6-0.1 \label{tab:SNREstimated}}
\tablehead{
\colhead{} & \multicolumn{2}{|c}{Hadronic} & \multicolumn{5}{|c}{Leptonic} \\
\colhead{$L_{\gamma}^{\mathrm{HAWC}}$ [erg/s]} & \multicolumn{1}{|c}{$n_{h,9.6{\mathrm{kpc}}}$ [cm${}^{-3}$]} & \colhead{$W_{p}$ [erg]} & \multicolumn{1}{|c}{$\rho_{0,e\left(9.6\mathrm{kpc}\right)}$ [1/eV]} & \colhead{$\Gamma_{e}$} & \colhead{$E_{c,e}$ [TeV]} & \colhead{$B$ [$\mu $G]} & \colhead{$W_{e}$ [erg]}
}
\renewcommand{\arraystretch}{1.4}
\startdata
$1.38^{+0.27}_{-0.17} \times ~ 10^{35}$ & ${46}$ & {$2.37^{+1.59}_{-0.73}\times10^{49}$} & $8.03^{+0.54}_{-0.52}\times10^{33}$ & $2.37^{+0.01}_{-0.01}$ & $92.20^{+6.23}_{-6.70}$ & $2.85^{+0.17}_{-0.16}$ & $2.60^{+0.54}_{-0.45} \times ~ 10^{50}$ \\
\enddata
\tablecomments{
The best-fit parameters for the proton spectrum are same as Table~\ref{tab:scenarios}. The best-fit parameters for the electron spectrum ($\rho_{0e\left(9.6\mathrm{kpc}\right)}$, $\Gamma_{e}$, $E_{c,e}$) have been refitted at 9.6~kpc to include the synchrotron emission spectrum.}
\end{deluxetable*}
}

For the modeling of SNR~G28.6-0.1, radio and infrared data are available as well as the $\gamma$-ray data. Hence, the Naima framework has computed both the synchrotron and inverse Compton scattering spectrum for the leptonic scenario \citep{SNRradio,SNRIR}.
Despite the near-infrared observations at \citet{SNRIR}, the near-infrared photons are not considered as seed photons for inverse Compton scattering because photons with energies higher than the CMB are suppressed via the Klein-Nishina effect \citep{KNsuppression}. 
With the additional data points, we are able to model and fit the synchrotron regime as well as the inverse Compton scattering regime for the leptonic scenario. Table~\ref{tab:SNREstimated} shows the best-fit values of the newly defined leptonic scenario. {Note that since we use radio data to fit the leptonic model for SNR~G28.6-0.1}, we have extended $E_{\mathrm{min},e}$ to 10~MeV. {Here, we assume $E_{\mathrm{max},e}=740$~TeV, which is calculated in \citet{LHAASOPSR, Emaxelectron} based on the maximum photon energy of LHAASO~J1843-0338.}

The obtained magnetic field strength of $2.85^{+0.17}_{-0.16}~\mu$G {from the Naima fit} is comparable to that of the average Galactic magnetic field strength of $2-3~\mu$G \citep{Longair}. The hadronic energy budget has been derived from the proton spectrum in Table~\ref{tab:scenarios} since the hadronic model has been fitted with only the $\gamma$-ray data.

\begin{figure*}[ht]
\fig{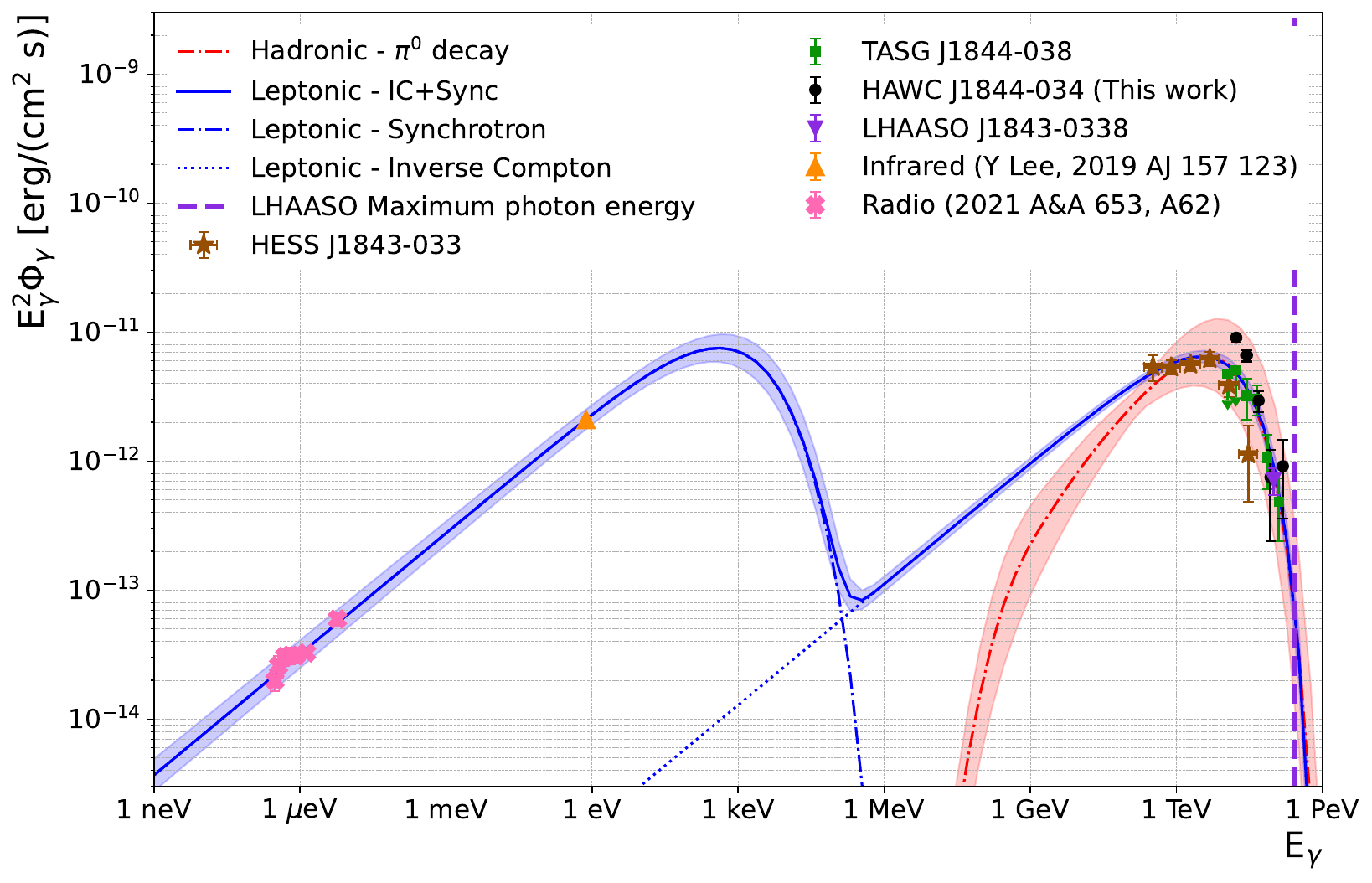}{2.0\columnwidth}{}
\caption{A spectral distributions plot with the best-fit leptonic (blue) and hadronic (red) scenarios for SNR~G28.6-0.1. \label{fig:SNRScenario}}
\end{figure*}

The photon spectra obtained from the best-fit scenarios of SNR~G28.6-0.1 are presented in Figure~\ref{fig:SNRScenario}. 
The gamma-ray data points are same as those of Figure~\ref{fig:TeVScenario}. The pink and orange data points are from radio and infrared observations, respectively \citep{SNRradio, SNRIR}, used for the modeling of synchrotron emission. 
The final leptonic and hadronic spectra for SNR~G28.6-0.1 as a counterpart to HAWC~J1844-034 is displayed as blue and red lines, respectively.
The leptonic spectrum (solid) consists of two sub-spectra, synchrotron emission (dash-dotted) and inverse Compton scattering (dotted). The synchrotron and the inverse Compton scattering spectra have been fitted simultaneously while assuming a single population of electrons is responsible for the production of both emissions. 

We have not used any X-ray data in the fit because the reported X-ray flux values for SNR~G28.6-0.1 from \citet{ASCA,Chandra2} do not contain any flux uncertainties. Since the X-ray band is located in the keV cutoff region of the synchrotron emission, having the X-ray data in the fit model can alter the spectral break parameter and directly influence the spectral break of the inverse Compton scattering as well. Therefore, incorporating reliable X-ray data should help us obtain a more complete understanding of SNR~G28.6-0.1 in the future.

The typical energy of a type Ia supernova explosion is expected to reach $\sim10^{51}$~erg \citep{SNRIa}. According to our fitted model, the energy budget for the leptonic scenario is $\sim30~\%$ of the typical explosion energy. As for the hadronic scenario, the energy budget is estimated to be {$\sim2$~\% of the explosion energy}, which is more than sufficient \citep{SNREfficiency}. Furthermore, from \citet{Tibet}, the diffusion timescale for protons can be estimated using,

\begin{eqnarray}\label{ProtonDiffusionTimescale}
    \tau_{\mathrm{diff}}\sim1.2\times10^{4}~\chi^{-1}\left(\frac{R_{\mathrm{sys}}}{20~\mathrm{pc}}\right)^{2} ~~~~~~~~~~\nonumber \\
    \times \left(\frac{E_{p}}{\mathrm{GeV}}\right)^{-0.5}\left(\frac{B}{10~\mu\mathrm{G}}\right)^{0.5}~\mathrm{yr},
\end{eqnarray}
where $\chi$ is the factor of diffusion suppression, $R_{\mathrm{sys}}$ is the radius of the emission region. 
For the calculation, $\chi$ of 0.1 and $B$ of $10~\mu$G are assumed \citep{Tibet}. Using $R_{\mathrm{sys}} = (0\fdg48~\pi d_{\mathrm{src}})/180\degr \sim 80$~pc and $E_{p} = 169$~TeV (cutoff energy from the best-fit proton spectrum), we can obtain $\tau_{\mathrm{diff}}$ of $\sim5$~kyr, which is indeed smaller than the known age of SNR~G28.6-0.1 of 19~kyr \citep{SNRAge}. 

\section{Conclusion} \label{subsec:Conclusion}

In this work, the previous eHWC~J1842-035 region has been thoroughly studied with improved dataset based on a systematic approach that precisely models the region. The best-fit model of the region includes three sources within the ROI. Of the three sources, HAWC~J1844-034 has been found to have the highest TS with the best-fit location of ($\alpha,~\delta$)=($281\fdg02^{+0.05}_{-0.05},~-3\fdg64^{+0.05}_{-0.04}$) and a best-fit extension of $0\fdg48^{+0.02}_{-0.02}$. The highest $\gamma$-ray photon energy of HAWC~J1844-034 is measured at 175~TeV with a differential flux $1.32^{+0.10}_{-0.09}\times10^{-14}$~(TeV~cm${}^{2}$~s)${}^{-1}$ at 20~TeV. By increasing the statistics from 1,039 days to 1,910 days and lowering the energy threshold from 56~TeV to 17.8~TeV, HAWC~J1844-034 could be studied in great depth.

We have confirmed that HAWC~J1844-034 has spectral and spatial coherence with the previous multi-TeV observations, LHAASO~J1843-0338 and TASG~J1844-038. The most-likely sub-TeV counterpart has been determined to be HESS~J1843-033, which has a similar flux magnitude to HAWC~J1844-034. There is no plausible GeV counterpart for HAWC~J1844-034 at this time.

We have fitted particle spectra based on two scenarios to find the origin of the $\gamma$-ray emission from HAWC~J1844-034. The hadronic scenario shows the proton cutoff region at $\sim169$~TeV, so classifying HAWC~J1844-034 as a PeVatron would be challenging.

Furthermore, to identify the nature of HAWC~J1844-034, we have examined four nearby pulsars and a supernova remnant. {We have found PSR~J1844-0346 as the most favorable candidate among the nearby pulsars since it is the only pulsar that satisfies both the spin-down power budget and having the pulsar contained within the extension of HAWC~J1844-034.}

SNR~G28.6-0.1 is another potential counterpart source that has spatial coherence with HAWC~J1844-034 {since it is also located within the extension of HAWC~J1844-034}. For the hadronic case of SNR~G28.6-0.1, the diffusion timescale of the protons is reasonable when compared with the estimated age of the SNR, and it has a plausible energy budget to be a counterpart of HAWC~J1844-034. {The leptonic scenario is also viable for SNR~G28.6-0.1. It would require $\sim30$~\% of the total explosion energy used as the acceleration of particles.}

\section{Acknowledgements} \label{subsec:Acknowloedgements}
We acknowledge the support from the US National Science Foundation (NSF); the US Department of Energy Office of High-Energy Physics; the Laboratory Directed Research and Development (LDRD) program of Los Alamos National Laboratory; Consejo Nacional de Ciencia y Tecnolog\'{\i}a (CONACyT), M{\'e}xico, grants 271051, 232656, 260378, 179588, 254964, 258865, 243290, 132197, A1-S-46288, A1-S-22784, c{\'a}tedras 873, 1563, 341, 323, Red HAWC, M{\'e}xico; DGAPA-UNAM grants IG101320, IN111315, IN111716-3, IN111419, IA102019, IN112218; VIEP-BUAP; PIFI 2012, 2013, PROFOCIE 2014, 2015; the University of Wisconsin Alumni Research Foundation; the Institute of Geophysics, Planetary Physics, and Signatures at Los Alamos National Laboratory; Polish Science Centre grant, DEC-2017/27/B/ST9/02272; Coordinaci{\'o}n de la Investigaci{\'o}n Cient\'{\i}fica de la Universidad Michoacana; Royal Society—Newton Advanced Fellowship 180385; Generalitat Valenciana, grant CIDEGENT/2018/034; Chulalongkorn Universitys CUniverse (CUAASC) grant; Instituto de F\'{\i}sica Corpuscular, Universitat de Val{\`e}ncia grant E-46980; and National Research Foundation of Korea grant NRF-2021R1A2C1094974. Thanks to Scott Delay, Luciano D\'{\i}az, and Eduardo Murrieta for technical support.

\bibliography{J1844}{}
\bibliographystyle{aasjournal}

\end{document}